# Generalized micropolar continualization of 1D beam lattices

Andrea Bacigalupo[1] and Luigi Gambarotta[2*]

[1]IMT School for Advanced Studies, Lucca, Italy
[2]Department of Civil, Chemical and Environmental Engineering, University of Genova, Italy

**Abstract**

The enhanced continualization approach proposed in this paper is aimed to overcome some drawbacks observed in the homogenization of beam lattices. To this end an enhanced homogenization technique is proposed and formulated to obtain consistent micropolar continuum models of the beam lattices and able to simulate with good approximation the boundary layer effects and the Floquet-Bloch spectrum of the Lagrangian model. The continualization technique here proposed is based on a transformation of the difference equation of motion of the discrete system via a proper down-scaling law into a pseudo-differential problem; a further McLaurin approximation is applied to obtain a higher order differential problem. The formulation is carried out for simple one-dimensional beam lattices that are, nevertheless, characterized by a rather wide variety of static and dynamic behaviors: the rod lattice, the beam lattice with node rotations and a 1D beam lattice model with generalized displacements. Higher order models may be obtained which are characterized by differential problems involving non-local inertia terms together with spatial high gradient terms. Moreover, the homogenized models obtained by the proposed enhanced continualization technique turn out to be energetically consistent and provide a good simulation of both the static response and of the acoustic spectrum of the original discrete models. The proposed homogenization procedure is first presented for the simple case of monoatomic axial chains. The beam lattice with node rotations and displacement prevented exhibits, in the static regime, decaying oscillations of the nodal rotation in the boundary layer which is well simulated by the homogenized model obtained by the proposed approach. Similar good results are obtained in the simulation of the optical spectrum. It is worth to note that in this case the homogenized model obtained via Padé approximation turns out to be energetically non-consistent. The analysis of the homogenized model derived from the beam lattice with transverse displacement and rotation of the nodes with elastic supports has shown that both the static and the dynamic response are strongly variable on the parameters of the Lagrangian model. Finally, several different cases have been considered and good simulations have been obtained both in describing the static response to prescribed displacements at the end nodes and in representing the Floquet-Bloch spectrum and the polarization vectors.

---

[*] Corresponding Author

# 1. Introduction

Since Newton's theory of sound propagation, lattices have constituted a class of fundamental models in many scientific fields. In Mechanics, lattice models are applied to capture, at the relevant scales of observation, the phenomena related to the discreteness of the systems to be represented, from the propagation of dispersive elastic waves (see Brillouin, 1946) to the stiffness and strength of inhomogeneous materials (see Fleck *et al.*, 2010), just to mention a few. Although the simplicity of lattice models allows in many cases to obtain analytical results, however, their application to systems with complex geometry and large dimensions involves a very high number of degrees of freedom and precludes the synthetic description of the results. Therefore, many studies have long been focused on the formulation of equivalent continua able to simulate the static and dynamic response of lattice, including the influence of size effects. To this end, techniques for the continualization of the discrete equations of the lattice motion have been proposed, which consist in replacing the discrete equations of motion and the nodal displacements in the lattice with differential equations of motion in terms of a suitably displacement field defined on an equivalent domain.

Higher order continuum models have been formulated to simulate the dispersive wave propagation occurring in one-dimensional chains (Brillouin, 1946), i.e. lattices with axial displacement of the nodes (see Askes *et al.*, 2002). These continuum models are derived through a continualization of the equations of motion of the Lagrangian model or equivalently of the energy functional, in which the difference of displacements of adjacent nodes is approximated by a truncated series of a macro-displacement field (Askes and Metrikine, 2005). Despite the simplicity of the approach, the homogenized models based on this standard continualization present loss of positivity of the elastic potential energy density, imaginary frequencies in the elastic field, unbounded group velocities in the short-wave limit, just to mention the main drawbacks (Metrikine and Askes, 2002).

In order to eliminate these pathologies several enhanced techniques have been proposed. The difference equations of motion of the Lagrangian model are transformed into a system of pseudo-differential equations which are approximated by the Padé approximant. As a consequence, non-local inertial terms, involving the spatial derivatives of the continuous acceleration field, are introduced indirectly in the continuum differential problem (Kevrekidis *et al.*, 2002, Rosenau, 2003, Andrianov and Awrejcewicz, 2008, Andrianov *et al.*, 2012, Challamel *et al.*, 2016, 2018). Alternatively, enhanced higher order models have been proposed relying on heuristic generalization of the down scaling law coupled with a perturbation method (Metrikine and Askes, 2002) or on the assumption that the PDE governing the dynamic behavior of the continuum model must be of the same order with respect to spatial coordinate and with respect to time (Metrikine, 2006). Different continualization approaches based on asymptotic schemes have been proposed for lattices having visco-elastic ligaments or non-linear behavior (see Reda *et al.*, 20016a, 2016b, 2017b, 2018)

More complex is the case of periodic beam lattices, whose nodes can undergo translations and rotations, even with rotational inertia, and are connected by ligaments having axial and bending stiffness. This involves a complex dynamic behavior due to the coupling of translational and rotational modes, with the existence of both acoustic branches and optical branches in the Floquet-Bloch spectrum. Suiker *et al.*, 2001, analyzed several 2D beam lattices and derived the equations of motion of the homogenized equivalent medium through a standard continualization, while Gonella and Ruzzene, 2008, and Lombardo and Askes, 2012, addressed the case in which the rotational inertia is neglected. An improved homogenization technique based on a multi-field approach has been formulated by Vasiliev *et al.*, 2008, 2010, 2014, and applied to square lattices endowed of rotational inertia. A different micropolar continualization including visco-elastic effects has been proposed by Reda et al., 2016c, 2017a,

It is important to note that the continualization of the difference equations of periodic beam lattices with a second order Taylor approximation of the generalized displacement field implies a non-positive definite potential energy density of the equivalent continuum. This result, already obtained by Bažant and Christensen, 1972, in developing a micropolar homogenization of large scale multistory framed structures, was taken up by Kumar and McDowell, 2004, addressing the problem of the static micropolar homogenization of plane beam-lattices. In particular, the energy density associated with the micro-curvatures turns out to be non-positive definite, a result that is also confirmed on the basis of energy equivalence. It is worth to note that other approaches based on a first order down-scaling law, in which the overall-elastic modules are obtained through the Hill-Mandel macro-homogeneity condition (see for example Pradel and Sab, 1998, Onck, 2002), involve a positive definite elastic energy density and therefore would seem not affected by the loss of ellipticity of the governing equations (see Kumar and McDowell, 2004). However, it has recently been shown by the authors (Bacigalupo and Gambarotta, 2017a) that the micropolar model obtained by the second-order standard continualization shows, despite the limits mentioned above, a remarkable accuracy in simulating the acoustic behavior of beam lattices. In fact, this model is able to catch the decreasing of the optical branch observed in the Lagrangian model with increasing the wave vector from the long-wavelength limit, a feature that is qualitatively opposite to that shown from the micropolar model with positive definite elastic energy density related to the micro-curvatures. It is worth highlighting that similar results were obtained in micropolar modeling of periodic materials made up of rigid blocks connected by elastic interfaces (Bacigalupo and Gambarotta, 2017), which provided some clarification about problems concerning waves propagation in periodic granular media (Merkel et. al., 2011).

These considerations emphasize the need for an enhanced homogenization technique to obtain thermodynamically consistent (in the sense of strong ellipticity) micropolar continuum models of the beam lattices able to simulate with good accuracy the boundary layer effects and Floquet-Bloch spectrum of the Lagrangian model, at least for long and medium wavelengths. In this paper an enhanced continualization technique is proposed that draws on the approach by Rosenau, 2003, and is formulated for simple one-dimensional beam lattices that are, nevertheless, characterized by a rather wide variety of static and dynamic behaviors. The most noticeable aspect is the presence of higher order terms both in the constitutive and in the inertial terms of the differential equation of motion of the equivalent continuum, which are derived independently. The proposed homogenization procedure is first presented for the simple case of monoatomic axial chains (Section 2). The difference equation of motion of the reference mass is transformed, through a suitable down-scaling law regularized to the first order, into a pseudo-differential equation and the differential equation of motion of the continuum model is obtained by approximating the pseudo-differential operators up to the required order term of its Taylor series. The capability of the resulting higher order continuum models to approximate the Lagrangian one is evaluated both in the static and in the dynamic field. Moreover, comparisons with the results from the models based on the standard continualization and on the Padé approximation are shown.

The second simple typology here considered consists of a continuous beam with rotating nodal masses, with prevented transverse displacement, connected by elastic ligaments modeled as Euler-Bernoulli beams (Section 3). This system exhibits a characteristic static response when rotations are imposed at the end nodes of a beam of finite length, with a spatially decaying of the nodal rotations and a decreasing optical branch in the acoustic spectrum. Also in this case the results of the enhanced continuum are compared with those of the Lagrangian model and with those of the continuum models derived through the standard continualization and the Padé approximation.

Finally, a third discrete system of continuous beam is considered equipped with equally spaced nodal masses undergoing transverse displacements and rotations, and elastic supports located at the nodes. This model may represent different discrete systems including rectangular lattices undergoing 1D generalized displacement fields (Section 4.a) and waveguides with passive control of the acoustic response (Section 4.b). Also for these systems both the static response to prescribed displacements at the end nodes and the acoustic spectrum are compared with those of the Lagrangian system and with those obtained in the continuum models derived through the standard continualization and the Padé approximation. The energetically consistent structure of the density of the elastic potential energy and of the kinetic energy of the continuous models derived with the proposed approach is discussed in detail, as well as the validity limits of the continuum models derived through the homogenization techniques considered. An extension of the proposed approach to 2D and 3D beam lattices will be presented by the Authors in a forthcoming paper.

## 2. 1D rod lattice model

Let us consider a 1-D lattice made up of equally spaced nodes with mass $m$ and connected through ligaments of length $\ell$ with axial stiffness $h$ (see Figure 1). The axial force applied on the $i$-th node is denoted by $f_i$ and $u_i$ is the displacement. In terms of the non-dimensional axial displacement $\psi_i = \dfrac{u_i}{\ell}$ the equation of motion if the $i$-th node is written as

$$\psi_{i-1} - 2\psi_i + \psi_{i+1} + \overline{f}_i = I_\psi \ddot{\psi}_i \,, \tag{1}$$

with $\overline{f}_i = \dfrac{f_i}{h\ell}$ and $I_\psi = \dfrac{m}{h}$. In case of prescribed displacements $\psi_0$ and $\psi_n$ at the end nodes 0 and $n$, respectively, the static problem with vanishing axial forces ($f_i = 0$) involves the equilibrium of all the nodes with a system of $n$-$1$ linear homogeneous equations whose solution is a linear interpolating function.

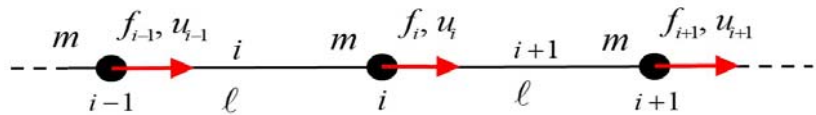

Figure 1. 1-D rod lattice: axial applied forces and nodal displacements.

In case of a lattice of free particles (see Brillouin, 1946), the harmonic axial wave propagation may be represented in the form $\psi_i(t) = \bar{\psi} \exp[j(kx_i - \omega t)]$, where $j^2 = -1$, $k$ is the wave number, $x_i$ is the location of the *i*-node with respect to the origin O. The angular frequency

$$\omega(k\ell) = \frac{2}{\sqrt{I_\psi}} \left|\sin\left(\frac{k\ell}{2}\right)\right| = \frac{k\ell}{\sqrt{I_\psi}} \left[1 - \frac{1}{24}(k\ell)^2 + \frac{1}{1920}(k\ell)^4\right] + \mathcal{O}(k^7 \ell^7), \qquad (2)$$

is a dispersion function with periodicity $2\pi$ and takes the role of acoustic branch in the Bloch spectrum.

It is well known that lattice models are inherently non-local. In fact, such models may exhibit boundary layer effects that occurs in static problems and their acoustic spectra are in general non-linear, namely presents dispersive wave propagations (see Eringen and Kim, 1977).

*Standard continualization*

To obtain an equivalent continuum model, a continualization approach is commonly applied that may be formulated by introducing the shift operator $E_\ell$ linking the displacement of two adjacent nodes, namely $\psi_{i+1} = E_\ell \, \psi_i$, for which the general property holds $\psi_{i \pm m} = (E_\ell)^{\pm m} \psi_i$, $m \in \mathbb{Z}^+$ (see Jordan, 1965, Rota et al., 1973, Kelley and Peterson, 2001, Andrianov and Awrejcewicz, 2008, Lombardo and Askes, 2010, for reference). Accordingly, the governing equation (1) may be written in the form

$$E_\ell \psi_i - 2\psi_i + (E_\ell)^{-1} \psi_i + \bar{f}_i = I_\psi \ddot{\psi}_i. \qquad (3)$$

where the shift operator is expressed as $E_\ell = \sum_{h=0}^{\infty} \frac{\ell^h}{h!} D^h = \exp(\ell D)$ with $D^h = \frac{\partial^h}{\partial x^h}$ (see Maslov, 1976, Shubin 1987). If a continuous field $\Psi(x,t)$ is introduced to represent the non-dimensional displacement in the equivalent continuum model with $\psi_i(t) = \Psi(x_i, t)$ and the nodal force is assumed as the resultant of a step-wise distribution $\bar{f}_i = \bar{f} \ell$, then equation (3) may be converted into the pseudo-differential equation

$$[\exp(\ell D) - 2 + \exp(-\ell D)] \Psi + \bar{f} = I_\psi \ddot{\Psi}, \qquad (4)$$

where the pseudo-differential operator $P(D) = [\exp(\ell D) - 2 + \exp(-\ell D)]$ may be identified and $\exp(\pm \ell D)\Psi(x) = \sum_{h=0}^{\infty} \frac{(\pm \ell)^h}{h!} \frac{\partial^h \Psi}{\partial x^h}$ (see Appendix A for details).

In order to obtain a differential equation of motion, the pseudo-differential operator may be expanded into McLaurin series

$$P(D) = \ell^2 D^2 + \frac{1}{12}\ell^4 D^4 + \frac{1}{360}\ell^6 D^6 + \mathcal{O}(\ell^7 D^7) . \tag{5}$$

When retaining the terms up to the second order, the classical equation of the elastic rod is obtained

$$\ell^2 \frac{\partial^2 \Psi}{\partial x^2} + \bar{f}\ell = I_\psi \ddot{\Psi} , \tag{6}$$

whose response to prescribed displacements at the rod ends is linear in agreement with the discrete model. However, it is well known that the wave propagation derived by equation (6) is non dispersive $\omega = \frac{k\ell}{\sqrt{I_\psi}}$, consequently this continuum model provides a frequency spectrum that agree with the discrete one only for the long wave-length limit $k\ell \to 0$. If a second order term is retained in expansion (5), a second-gradient equivalent continuum is obtained having the following governing equation

$$\ell^2 \frac{\partial^2 \Psi}{\partial x^2} + \frac{1}{12}\ell^4 \frac{\partial^4 \Psi}{\partial x^4} + \bar{f}\ell = I_\psi \ddot{\Psi} , \tag{7}$$

that is the Euler-Lagrange equation derived by assuming the following Lagrangian density function

$$\mathcal{L} = \frac{1}{2\ell} I_\psi \dot{\Psi}^2 - \frac{1}{2\ell}\left[\ell^2 \left(\frac{\partial \Psi}{\partial x}\right)^2 - \frac{\ell^4}{12}\left(\frac{\partial^2 \Psi}{\partial x^2}\right)^2\right] + \bar{f}\Psi , \tag{8}$$

with not-positive-definite elastic potential density (see Challamel *et al.*, 2016). The wave propagation in model (7) is dispersive with dispersion function

$$\omega = \frac{k\ell}{\sqrt{I_\psi}}\sqrt{\left(1 - \frac{1}{12}k^2\ell^2\right)} = \frac{k\ell}{\sqrt{I_\psi}}\left[1 - \frac{1}{24}k^2\ell^2 - \frac{1}{1152}k^4\ell^4\right] + \mathcal{O}(k^7\ell^7) . \tag{9}$$

As it is well known, the second gradient model turns out to be rather accurate in describing the dispersion function in the neighbor of the long-wave limit. Moreover, it must be emphasized that for $k\ell > 2\sqrt{3}$ the dispersion equation (9) loses a physical meaning and the Legendre-Hadamard ellipticity condition, requiring real values for the wave velocity, turns out to be not satisfied. Moreover, the non-positivity definiteness of the elastic potential energy density prevents the application of this model in static problems. If the next higher-order term is retained in (5) the dispersion function turns out to be positive, but its application is limited since in the short-wave limit $(k\ell \to \infty)$ the group velocity is unbounded (see Metrikine and Askes, 2002). It is worth to note that the same results are obtained if the continualization of the energy functional is applied, as shown by Askes and Metrikine, 2005.

*Enhanced continualization via first order regularization approach*

A different approach is here considered to obtain a continuum model able to simulate both the static and the dynamic response of the discrete Lagrangian system in terms of a macroscopic displacement field $\Psi(x,t)$. The derivative at $x_i$ of such function is assumed to be related to the central difference as follows

$$\left.\frac{\partial \Psi}{\partial x}\right|_{x_i} \doteq \frac{\psi_{i+1} - \psi_{i-1}}{2\ell} . \tag{10}$$

By introducing the shift operator previously defined, the derivatives of the continuum fields may be expressed as

$$\left.\frac{\partial \Psi}{\partial x}\right|_{x_i} = D\Psi\big|_{x_i} = \frac{\exp(\ell D) - \exp(-\ell D)}{2\ell} \psi_i , \tag{11}$$

from which one obtains the down-scaling law for the node translation in terms of the continuous field

$$\psi_i(t) = \frac{2\ell D}{\exp(\ell D) - \exp(-\ell D)} \Psi(x,t)\bigg|_{x_i} , \tag{12}$$

see Appendix B for details.

By substituting equation (12) into the discrete equations of motion (1) and applying the shift operator, the pseudo-differential equation of motion of the equivalent continuum takes the form

$$\left\{ \frac{2\left[\exp(\ell D) - 2 + \exp(-\ell D)\right]}{\left[\exp(\ell D) - \exp(-\ell D)\right]} \ell D \right\} \Psi + \bar{f} = \left\{ \frac{2I_\psi}{\left[\exp(\ell D) - \exp(-\ell D)\right]} \ell D \right\} \ddot{\Psi} , \quad (13)$$

where the pseudo-differential operators $P_1(D) = \dfrac{2\left[\exp(\ell D) - 2 + \exp(-\ell D)\right]}{\left[\exp(\ell D) - \exp(-\ell D)\right]} \ell D$ and

$P_2(D) = \dfrac{2\ell D}{\left[\exp(\ell D) - \exp(-\ell D)\right]}$ may be identified, whose McLaurin expansions are

$$\begin{aligned}
P_1(D) &= \ell^2 D^2 - \frac{1}{12}\ell^4 D^4 + \frac{1}{120}\ell^6 D^6 + \mathcal{O}(\ell^7 D^7) , \\
P_2(D) &= 1 - \frac{1}{6}\ell^2 D^2 + \frac{7}{360}\ell^4 D^4 - \frac{31}{15120}\ell^6 D^6 + \mathcal{O}(\ell^7 D^7) .
\end{aligned} \quad (14)$$

When retaining the terms up to the second order, the equation of motion of the equivalent continuum takes the form

$$\ell^2 \frac{\partial^2 \Psi}{\partial x^2} + \bar{f}\ell = I_\psi \left( \ddot{\Psi} - \frac{\ell^2}{6} \frac{\partial^2 \ddot{\Psi}}{\partial x^2} \right) , \quad (15)$$

corresponding, in the static range, to equation (6) (obtained via a first order standard continualization) and (20) (derived as a first order continualization via Padé approximant), while in the dynamic range a non-local inertia term appears. The dispersion function takes the form

$$\omega = k\ell \sqrt{\frac{1}{I_\psi \left(1 + \frac{1}{6}k^2\ell^2\right)}} = \frac{k\ell}{\sqrt{I_\psi}} \left( 1 - \frac{1}{12}k^2\ell^2 + \frac{1}{96}k^4\ell^4 \right) + \mathcal{O}(k^5\ell^5) . \quad (16)$$

If terms up to the fourth order are retained in (14), the equation of motion of the homogenized continuum takes the form

$$\ell^2 \frac{\partial^2 \Psi}{\partial x^2} - \frac{\ell^4}{12} \frac{\partial^4 \Psi}{\partial x^4} + \bar{f}\ell = I_\psi \left( \ddot{\Psi} - \frac{\ell^2}{6} \frac{\partial^2 \ddot{\Psi}}{\partial x^2} + \frac{7\ell^4}{360} \frac{\partial^4 \ddot{\Psi}}{\partial x^4} \right) . \quad (17)$$

In this case the dispersion function is

$$\omega = \frac{k\ell}{\sqrt{I_\psi}} \sqrt{\frac{1 + \frac{1}{12}k^2\ell^2}{1 + \frac{k^2\ell^2}{6} + \frac{7}{360}k^4\ell^4}} = \frac{k\ell}{\sqrt{I_\psi}} \left( 1 - \frac{1}{24}k^2\ell^2 - \frac{7}{1920}k^4\ell^4 \right) + \mathcal{O}(k^5\ell^5) . \quad (18)$$

It is worth to note that equation (17) is the Euler-Lagrangian equation derived by assuming the following Lagrangian density function

$$\mathcal{L} = \frac{1}{2\ell} I_\psi \left[ \dot{\Psi}^2 + \frac{\ell^2}{6} \left(\frac{\partial \dot{\Psi}}{\partial x}\right)^2 + \frac{7\ell^4}{360} \left(\frac{\partial^2 \dot{\Psi}}{\partial x^2}\right)^2 \right] - \frac{1}{2\ell} \left[ \ell^2 \left(\frac{\partial \Psi}{\partial x}\right)^2 + \frac{\ell^4}{12} \left(\frac{\partial^2 \Psi}{\partial x^2}\right)^2 \right] + \bar{f}\Psi \,, \quad (19)$$

with both the elastic potential density energy and the kinetic density energy positive definite. It is worth to note that this approach allows to introduce non-local inertial terms from the series approximation of the inertial pseudo-differential operator. These terms do not depend on those deriving from the approximation of the pseudo-differential constitutive operator, unlike what happens in the approach based on the Padé approximant (see Section 3). A different continualization model in non-local integral continua is derived in Appendix C, which is based on two-sided Zeta transform applied to the discrete governing equation.

*Continualization via Padé approximant*

It is worth to note that nonlocal inertia terms may be obtained on the basis of Padé approximations (see for reference Kevrekidis *et al.*, 2002, and Cuyt, 1980). In this case the equation of motion is given in the form

$$\ell^2 \frac{\partial^2 \Psi}{\partial x^2} + \bar{f}\ell = I_\psi \left( \ddot{\Psi} - \frac{\ell^2}{12} \frac{\partial^2 \ddot{\Psi}}{\partial x^2} \right) \,, \quad (20)$$

and the homogenized continuum is characterized by positive definite elastic potential energy density and kinetic energy density; the dispersion function takes the form

$$\omega = k\ell \sqrt{\frac{1}{I_\psi \left(1 + \frac{1}{12} k^2 \ell^2\right)}} = \frac{k\ell}{\sqrt{I_\psi}} \left(1 - \frac{1}{24} k^2 \ell^2 + \frac{1}{384} k^4 \ell^4\right) + \mathcal{O}(k^5 \ell^5). \quad (21)$$

*Benchmark test for the continualization approaches*

A comparison of the dispersion functions in the Bloch spectrum obtained by the considered homogenization approaches with the dispersion function by the Lagrangian model is shown in the diagrams of Figure 2. In the diagrams of Figure 2.a it may be observed that the best approximation to the exact solution (black line) is obtained through the fourth order proposed model (red line) also in the short wavelength regime $(k\ell \approx \pi)$. However, also the second order proposed model (violet line) seems to be in good agreement also in the short wavelength regime, namely for wavelength $\lambda \approx 2\ell$; a capability that is not shown by the continuum model obtained via Padé approximant (yellow line). On the contrary, the dispersion function obtained through the standard continualization (blue line) does not provide good accuracy in the short wavelength regime. It is worth to note that the considered models, with the exception of the models based on the standard continualization, are characterized by a positive definite elastic potential energy density and kinetic energy density. In the diagrams of Figure 2.b, the dispersion function of the Lagrangian model (black line) is compared with the dispersion functions by the enhanced homogenized model obtained for different orders of approximation.

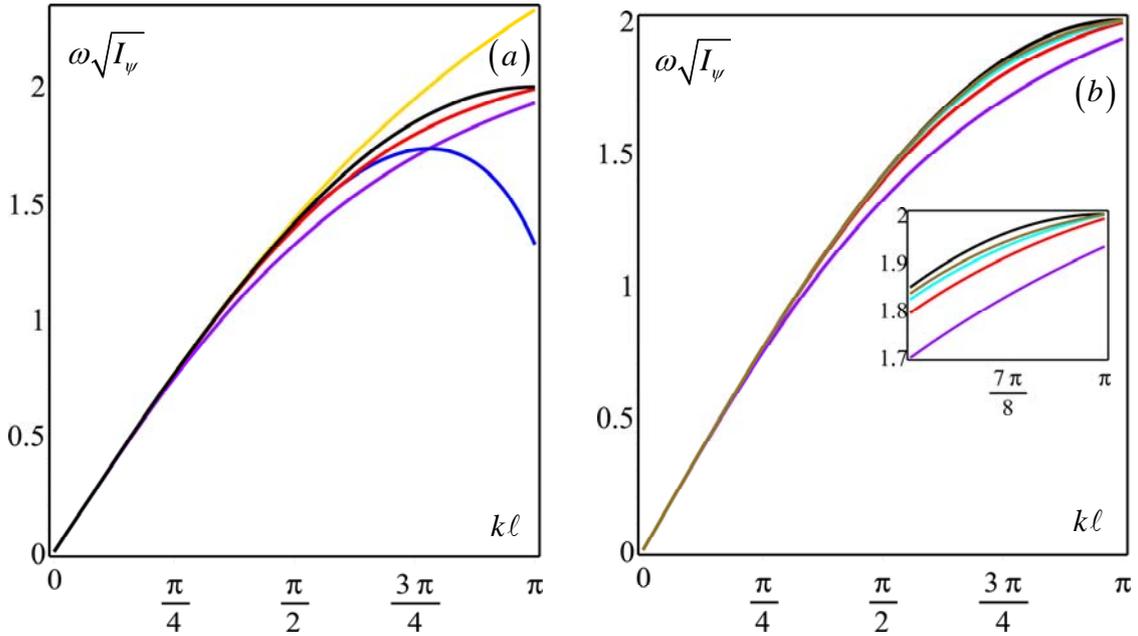

Figure 2. Dispersion functions of the 1D rod lattice. (a) Comparison of different models with the Lagrangian model (black line- equation (2)); 4$^{th}$ order standard continualization (blue – equation (9)); proposed 2$^{nd}$ order continualization (violet – equation (16)); proposed 4$^{th}$ order continualization (red – equation (18)); 2$^{nd}$ order continualization via Padé approximant (yellow – equation (21)). (b) Sensitivity of the proposed model on the considered order of approximation: 6$^{th}$ order continualization (cyan); 8$^{th}$ order continualization (brown).

## 3. 1D beam lattice with node rotations

Let now consider a lattice made of equally spaced nodes free to rotate in the same plane with restrained displacements. The nodes have equal rotational inertia $I$ and are connected through elastic Euler-Bernoulli beams as shown in figure 3. The deformed configuration of the lattice is defined here, by the nodal rotations $\varphi_i$ of the $i$-th node to which the applied couple $c_i$ is energetically associated. The kinetic energy of the $i$-th nodal mass is $T_i = \frac{1}{2} I \dot{\varphi}_i^2$. The elastic potential energy of the $i$-th ligament between the $i$-th node and the $i+1$ node is $\Pi_{ei} = \frac{1}{2} 4H \left( \varphi_i^2 + \varphi_i \varphi_{i-1} + \varphi_{i-1}^2 \right)$, being $H = \frac{EJ}{\ell}$ the flexural stiffness. Moreover, in case of applied conservative generalized forces, the corresponding potential energy is $\Pi_{fi} = -c_i \varphi_i$. The Euler-Lagrange equation of motion of the $i$-th node is written in the form

$$-\frac{1}{6}\left(\varphi_{i-1} + 4\varphi_i + \varphi_{i+1}\right) + \overline{c}_i = I_\varphi \ddot{\varphi}_i \ , \qquad (22)$$

being $I_\varphi = \frac{I}{12H}$ and $\overline{c}_i = \frac{c_i}{12H}$. It is worth noting that the model corresponds to a continuous beam loaded at the supports by couples and with possible rotations prescribed at the end nodes.

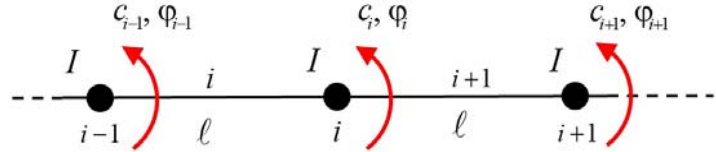

Figure 3. 1-D beam lattice: applied couples and nodal rotations.

The solution of the static linear difference equation of the finite system having $n+1$ nodes with vanishing nodal couples ($c_i = 0$, $i = 0, n$) and prescribed rotations $\varphi_0$ and $\varphi_n$ at the end nodes is carried out according to Kelley and Peterson, 2001, and takes the form

$$\varphi_i = C_1 (-1)^i \left(2 - \sqrt{3}\right)^i + C_2 (-1)^i \left(2 + \sqrt{3}\right)^i \ . \qquad (23)$$

This solution exhibits growing or decaying oscillation of the nodal rotations as an effect of the second term in (23) depending on $(-1)^i$ and where the two constants depend on the boundary conditions as follows

$$C_1 = \frac{\varphi_n - (-1)^n (2+\sqrt{3})^n \varphi_0}{(-1)^n \left[(2-\sqrt{3})^n - (2+\sqrt{3})^n\right]} \qquad C_2 = \frac{\varphi_n - (-1)^n (2-\sqrt{3})^n \varphi_0}{(-1)^n \left[(2-\sqrt{3})^n - (2+\sqrt{3})^n\right]} , \qquad (24)$$

where the dependence of the constants on the number of nodes $n$ appears. Moreover, in comparison to the linear solution of the discrete problem of Section 2, the solution (23) presents an exponential type structure describing boundary layers.

The propagation of the rotational waves turns out to be dispersive and the dispersion function takes the form

$$\omega = \sqrt{\frac{2+\cos(k\ell)}{3 I_\varphi}} = \frac{1}{\sqrt{I_\varphi}} \left(1 - \frac{1}{12} k^2 \ell^2 + \frac{1}{288} k^4 \ell^4\right) + \mathcal{O}(k^6 \ell^6) , \qquad (25)$$

and represents an optical branch in the Bloch spectrum with periodicity $2\pi$.

*Standard continualization*

In developing a standard continualization, the function $\Phi(x,t)$ is introduced to represent the rotational field in the equivalent continuum. Once introduced the shift operator, a pseudo-differential equation is derived

$$-\frac{1}{6}\left[\exp(-\ell D) + 4 + \exp(\ell D)\right] \Phi + \bar{c}\ell = I_\varphi \ddot{\Phi} , \qquad (26)$$

with the nodal couple assumed as the resultant of a step-wise distribution $\bar{c}_i = \bar{c}\ell$. If the corresponding pseudo-differential operator is expanded in $D$ in analogy with (5) and the terms up to the second order are retained, the differential equation of motion is obtained

$$-\frac{1}{6}\ell^2 \frac{\partial^2 \Phi}{\partial x^2} - \Phi + \bar{c}\ell = I_\varphi \ddot{\Phi} . \qquad (27)$$

The corresponding Lagrangian density function is

$$\mathcal{L} = \frac{1}{2} I_\varphi \dot{\Phi}^2 - \frac{1}{2}\left[\Phi^2 - \frac{\ell^2}{6}\left(\frac{\partial \Phi}{\partial x}\right)^2\right] + \bar{c}\Phi , \qquad (28)$$

with non-positive-definite elastic energy density. When considering the static homogeneous problem with rotations prescribed at the end nodes, equation (27) takes the form of a 1D Helmholtz equation whose solution is

$$\Phi(x,t) = C_1 \sin\left(\frac{\sqrt{6}x}{\ell}\right) + C_2 \cos\left(\frac{\sqrt{6}x}{\ell}\right), \tag{29}$$

characterized by an oscillating behaviour depending on the cell size $\ell$ and on the macroscopic structural size through the constants $C_1$ and $C_2$ (which differ from those in the solution (23) related to the Lagrangian model). The dispersion function is

$$\omega = \sqrt{\frac{1}{I_\varphi}\left(1 - \frac{1}{6}k^2\ell^2\right)} = \frac{1}{\sqrt{I_\varphi}}\left(1 - \frac{1}{12}k^2\ell^2 - \frac{1}{288}k^4\ell^4\right) + \mathcal{O}(k^6\ell^6), \tag{30}$$

with a critical point in the Bloch spectrum in the longwave limit, where the group velocity vanishes $v_g(k\ell \to 0) = \left.\frac{d\omega}{dk}\right|_{k\ell\to 0} = 0$ and from where an optical branch departs.

Retaining a further term in the McLaurin expansion, the following higher order continuum model is obtained

$$-\Phi - \frac{1}{6}\ell^2 \frac{\partial^2 \Phi}{\partial x^2} - \frac{1}{72}\ell^4 \frac{\partial^4 \Phi}{\partial x^4} + \overline{c}\ell = I_\varphi \ddot{\Phi}, \tag{31}$$

to which a pathological Lagrangian density function is associated, with non-positive-definite elastic potential energy density

$$\mathcal{L} = \frac{1}{2\ell}I_\varphi \dot{\Phi}^2 - \frac{1}{2\ell}\left[\Phi^2 - \frac{\ell^2}{6}\left(\frac{\partial \Phi}{\partial x}\right)^2 + \frac{\ell^4}{72}\left(\frac{\partial^2 \Phi}{\partial x^2}\right)^2\right] + \frac{1}{\ell}\overline{c}\Phi. \tag{32}$$

In this case the optical dispersion function is obtained

$$\omega = \sqrt{\frac{1}{I_\varphi}\left(1 - \frac{1}{6}k^2\ell^2 + \frac{1}{72}k^4\ell^4\right)}\omega = \frac{1}{\sqrt{I_\varphi}}\left(1 - \frac{1}{12}k^2\ell^2 + \frac{1}{288}k^4\ell^4\right) + \mathcal{O}(k^6\ell^6), \tag{33}$$

with the same critical point at the longwave limit.

*Enhanced continualization via first order regularization approach*

If the downscaling law to represent the rotation of the nodes in terms of a macro-rotation field $\Phi(x,t)$ is assumed in analogy to the procedure presented in equations (10), (11) and (12), the equation of motion (22) takes the form of a pseudo-differential problem

$$\left\{-\frac{[\exp(\ell D)+4+\exp(-\ell D)]}{3[\exp(\ell D)-\exp(-\ell D)]}\ell D\right\}\Phi+\overline{c}=\left\{\frac{2I_\psi}{[\exp(\ell D)-\exp(-\ell D)]}\ell D\right\}\ddot{\Phi}. \qquad (34)$$

The terms in the braces may be expanded into a McLaurin series; the expansion truncated up to the fourth order provides the differential equation of motion of the equivalent continuum system

$$-\Phi-\frac{\ell^4}{180}\frac{\partial^4\Phi}{\partial x^4}+\overline{c}\ell=I_\varphi\left[\ddot{\Phi}-\frac{\ell^2}{6}\frac{\partial^2\ddot{\Phi}}{\partial x^2}+\frac{7\ell^4}{360}\frac{\partial^4\ddot{\Phi}}{\partial x^4}\right], \qquad (35)$$

to which the Lagrangian density function is associated

$$\mathcal{L}=\frac{1}{2\ell}I_\varphi\left(\dot{\Phi}^2+\frac{\ell^2}{6}\left(\frac{\partial\dot{\Phi}}{\partial x}\right)^2+\frac{7\ell^4}{360}\left(\frac{\partial^2\dot{\Phi}}{\partial x^2}\right)^2\right)-\frac{1}{2\ell}\left[\Phi^2+\frac{\ell^4}{180}\left(\frac{\partial^2\Phi}{\partial x^2}\right)^2\right]+\overline{c}\Phi, \qquad (36)$$

with both the elastic potential energy density and kinetic energy positive definite. The solution of the homogeneous problem with prescribed rotations at the ends takes the form

$$\Phi(x,t)=\left(C_1 e^{-\sqrt[4]{45}\frac{x}{\ell}}+C_2 e^{\sqrt[4]{45}\frac{x}{\ell}}\right)\sin\left(\sqrt[4]{45}\frac{x}{\ell}\right)+\left(C_3 e^{-\sqrt[4]{45}\frac{x}{\ell}}+C_4 e^{\sqrt[4]{45}\frac{x}{\ell}}\right)\cos\left(\sqrt[4]{45}\frac{x}{\ell}\right), \qquad (37)$$

and the dispersion function is written as

$$\omega=\sqrt{\frac{360+2k^4\ell^4}{I_\varphi(360+60k^2\ell^2+7k^4\ell^4)}}=\frac{1}{\sqrt{I_\varphi}}\left(1-\frac{1}{12}k^2\ell^2+\frac{1}{288}k^4\ell^4\right)+\mathcal{O}(k^6\ell^6). \qquad (38)$$

*Continualization via Padé approximant*

Finally, if the Padé approximant of the l.h.s of the pseudo-differential equation (26) is considered, namely

$$-\frac{1}{6}\left[\exp(-\ell D)+4+\exp(\ell D)\right]\Phi=-\frac{1}{3}\left[2+\cosh(\ell D)\right]\Phi\approx-\frac{1+\frac{1}{12}(\ell D)^2}{1-\frac{1}{12}(\ell D)^2}\Phi, \qquad (39)$$

the resulting differential equation of motion of the equivalent homogenized continuum takes the form

$$-\frac{\ell^2}{12}\frac{\partial^2\Phi}{\partial x^2}-\Phi+\overline{c}\ell=I_\varphi\left(\ddot{\Phi}-\frac{\ell^2}{12}\frac{\partial^2\ddot{\Phi}}{\partial x^2}\right). \qquad (40)$$

It is worth to note that the Lagrangian density function from which equation (40) is derived is

$$\mathcal{L}=\frac{1}{2\ell}I_{\varphi}\left[\dot{\Phi}^2+\frac{\ell^2}{12}\left(\frac{\partial\dot{\Phi}}{\partial x}\right)^2\right]-\frac{1}{2\ell}\left[\Phi^2-\frac{\ell^2}{12}\left(\frac{\partial\Phi}{\partial x}\right)^2\right]+\overline{c}\Phi \ . \tag{41}$$

Despite of the introduction of the Padé approximant, proposed in the literature to circumvent the loss of positive definiteness of the elastic potential energy density, in this case the goal is not achieved, unlike what is obtained with the proposed enhanced continualization (see equation (36)). As a consequence, also for this model an oscillating behavior with constant amplitude is obtained

$$\Phi(x,t)=C_1\sin\left(2\frac{\sqrt{3}x}{\ell}\right)+C_2\cos\left(2\frac{\sqrt{3}x}{\ell}\right) , \tag{42}$$

while the dispersion function is

$$\omega=\sqrt{\frac{1-\frac{1}{12}k^2\ell^2}{I_{\varphi}\left(1+\frac{1}{12}k^2\ell^2\right)}}=\frac{1}{\sqrt{I_{\varphi}}}\left(1-\frac{1}{12}k^2\ell^2+\frac{1}{288}k^4\ell^4\right)+\mathcal{O}\left(k^6\ell^6\right) \ . \tag{43}$$

The fourth order continuum model derived by a Padé approximation is given in Appendix D, where it is shown that a non-positive-definite elastic potential energy density is obtained.

*Benchmark tests for the continualization approaches*

The capabilities of the proposed homogenization approach may be assessed by the simulation of the discrete system in both the static and the dynamic field. As a static case, let us consider a continuous beam made up of $n=10$ ligaments subjected to a prescribed rotation $\varphi_0=10^{-2}$ at the end left node with restrained rotation at the right one $\varphi_{10}=0$. The rotations at the nodes of the discrete system given by (23) and (24) are represented in the diagram of Figure 4.a together with the macro-rotation field $\Phi(x)$ obtained by solving the governing differential equation derived by the proposed enhanced homogenization, via $r^{\text{th}}$ order continualization, with appropriate boundary conditions, i.e. $\Phi(x=0)=1$, $\Phi(x=n\ell)=0$ and $\left.\frac{d^h\Phi}{dx^h}\right|_{x=0,n\ell}=0$, $h=1,..,r/2$. Here different solutions are plotted which are obtained by truncating the terms in braces in the pseudo-differential equation (34) at increasing orders, starting from the fourth order (see equation (37)), while the solutions by the standard continualization and the Padé approximation have been ignored being characterized by a non-positive definite elastic potential energy. In Figure 4a the decaying oscillation of the nodal rotations of the discrete model close to the boundary layer is shown, which is simulated rather well by the enhanced continuum models here proposed, with a tendency to converge to the actual solution when increasing the order of the continuum model. On the contrary, the rotation field (29) obtained through the standard continualization, being represented by harmonic functions, does not agree with the nodal rotations provided by the discrete model.

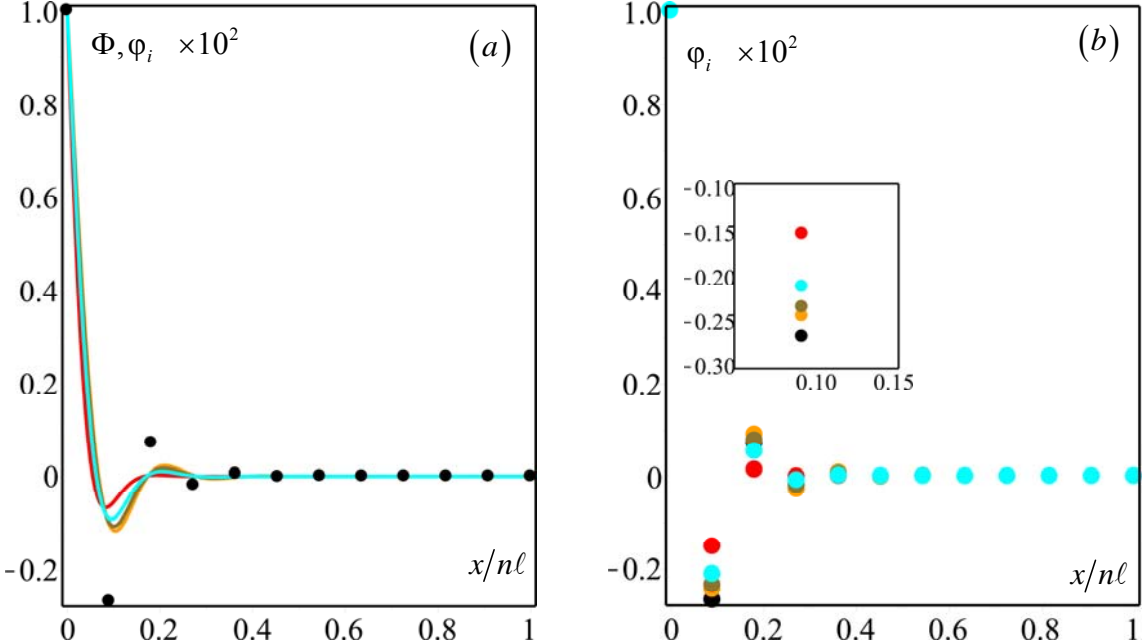

Figure 4. Node rotations in the Lagrangian model (black dots) vs. (a) rotational field in the enhanced continuum and (b) node rotations obtained via down-scaling relations by enhanced continuum: 4$^{th}$ order continualization (red); 6$^{th}$ order continualization (cyan); 8$^{th}$ order continualization (brown); 10$^{th}$ order continualization (orange); proposed 10$^{th}$ order continualization (orange).

The good accuracy of the enhanced homogenization here proposed may be observed in Figure 4b, where the nodal rotations from the Lagrangian model (black point) are compared with the nodal rotations obtained via downscaling

$$\varphi_i = \frac{2\ell D}{\exp(\ell D) - \exp(-\ell D)} \Phi(x)\bigg|_{x_i} =$$
$$= \left(1 - \frac{1}{6}\ell^2 \frac{d^2\Phi}{dx^2} + \frac{7}{360}\ell^4 \frac{d^4\Phi}{dx^4} - \frac{31}{15120}\ell^6 \frac{d^6\Phi}{dx^6}\right)\Psi(x_i) + \mathcal{O}(\ell^7) \quad (44)$$

from the enhanced models at the different orders of accuracy. Here, for the continuous model obtained via $r^{th}$ order continualization, the down-scaling relation has been carried out numerically by assuming estimates via a truncation at $r$-2 order.

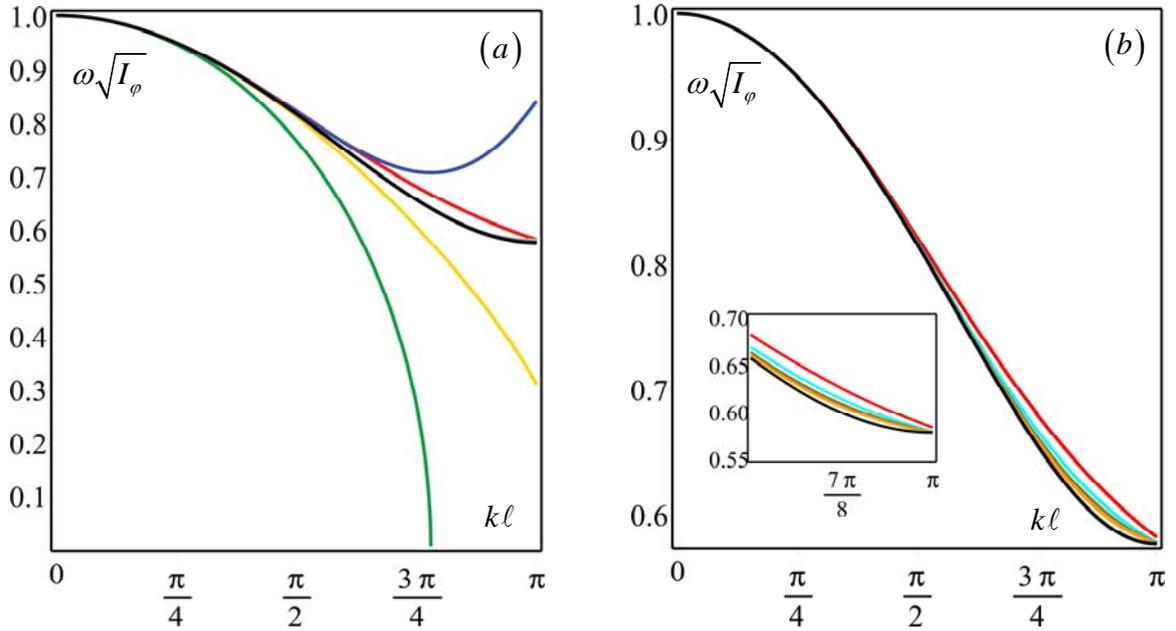

Figure 5. Dispersion functions of the 1D beam lattice with node rotation. (a) Comparison of different models: Lagrangian model (black line – equation (25)); standard continuous model $2^{nd}$ order continualization (green – equation (30)); standard continuous model $4^{th}$ order continualization (blue – equation (33)); proposed $4^{th}$ order continualization (red – equation (38)); $2^{nd}$ order continualization based on Padé approximation (yellow – equation (43)), $4^{th}$ order continualization based on Padé approximation (gray – equation (D.5)). Note that the black and the gray line are almost superimposed. (b) Sensitivity of the proposed model on the considered order of approximation: $6^{th}$ order continualization (cyan); $8^{th}$ order continualization (brown); $10^{th}$ order continualization (orange).

The validity limits of the enhanced model to simulate the propagation of harmonic waves in the Lagrangian system may be appreciated in the Bloch spectrum of Figure 5.a. From these diagrams it appears that the best simulation is obtained by the fourth order model based on the Padé approximation (the grey line is almost superimposed on the black line), even if, as shown in Appendix D, the elastic potential density of this model is non-positive definite. On the other side, it is worth to note that the dispersion function (38) by the fourth order enhanced continuum model (red line) turns out to provide a good simulation of the dispersion function of the discrete model and is characterized by a positive definite elastic potential energy. Finally, in Figure 5.b the optical branch from the Lagrangian model is compared with those obtained through the enhanced continualization up to the $10^{th}$ order. It results that increasing the order of the enhanced continuum, the corresponding dispersion functions tend to the corresponding optical branch of the Lagrangian model.

## 4. 1D beam lattice model

Let consider now the more general 1D beam lattice shown in Figure 6. This model is derived from the one analyzed in the previous Section with transverse displacement and rotation elastically restrained with translational and rotational elastic restraints of stiffness $K_\psi$ and $K_\varphi$, respectively. The deformed configuration of the lattice is defined by the transverse deflection $v_i$ and the rotation $\varphi_i$ of the nodes; axial displacements are ignored. Transverse forces $f_i$ and couples $c_i$ are applied at the nodes. This model may be representative of different elastic systems, among which those dealt with in the next subsections. When assuming $K_\psi = K_\varphi = 0$ this model appears to be similar to those one analysed by Vasiliev *et al.*, 2010.

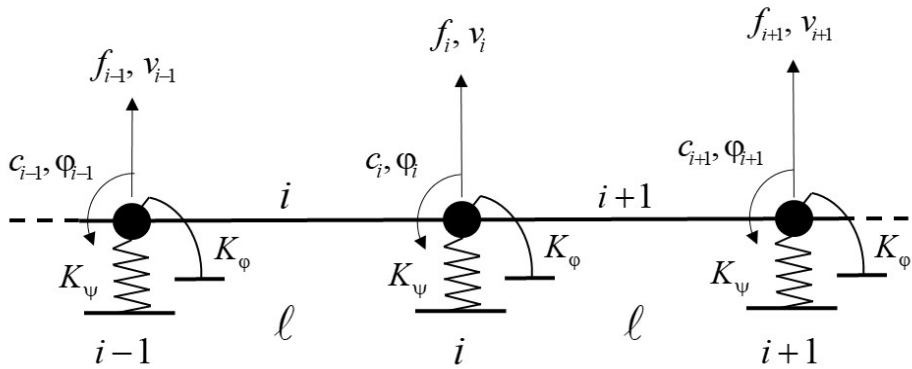

Figure 6. 1-D beam lattice.

The non-dimensional deflection is introduced $\psi_i = \dfrac{v_i}{\ell}$ and the kinetic energy of the *i*-th nodal mass is $T_i = \dfrac{1}{2}mv_i^2 + \dfrac{1}{2}I\varphi_i^2 = \dfrac{1}{2}m\ell^2\psi_i^2 + \dfrac{1}{2}I\varphi_i^2$. The elastic potential energy of the *i*-th ligament between node $i-1$ and node $i$ node is

$$\Pi_{ei} = \frac{1}{2} 4H \left[ \left( \varphi_i - \Delta\psi_i \right)^2 + \left( \varphi_i - \Delta\psi_i \right)\left( \varphi_{i-1} - \Delta\psi_i \right) + \left( \varphi_{i-1} - \Delta\psi_i \right)^2 \right], \quad \text{being} \quad H = \frac{EJ}{\ell} \quad \text{the}$$

flexural stiffness and $\Delta\psi_i = \psi_i - \psi_{i-1}$ the average rotation of the *i*-th ligament. The equivalent form $\Pi_{ei} = \frac{H}{2} \left\{ 12 \left[ \psi_i - \psi_{i-1} - \frac{1}{2}(\varphi_i + \varphi_{i-1}) \right]^2 + (\varphi_i - \varphi_{i-1})^2 \right\}$ may be considered that is in agreement to Vasiliev *et al.*, 2008. The elastic potential energy stored in the elastic restraints at the *i*-th node is $\Pi_{Ri} = \frac{1}{2}\left( K_\psi \ell^2 \psi_i^2 + K_\varphi \varphi_i^2 \right)$. Moreover, in case of applied conservative generalized forces, the corresponding potential energy is $\Pi_{fi} = -f_i \psi_i \ell - c_i \varphi_i$. The resulting Euler-Lagrange equations of motion of the *i*-th node are written as a system of difference linear equations

$$\begin{cases} \psi_{i-1} - \left( 2 + \tilde{K}_\psi \right)\psi_i + \psi_{i+1} - \frac{1}{2}\left( \varphi_{i+1} - \varphi_{i-1} \right) + \overline{f}_i = I_\psi \ddot{\psi}_i \\ \frac{1}{2}\left( \psi_{i+1} - \psi_{i-1} \right) - \frac{1}{6}\left[ \varphi_{i-1} + \left( 4 + 6\tilde{K}_\varphi \right)\varphi_i + \varphi_{i+1} \right] + \overline{c}_i = I_\varphi \ddot{\varphi}_i \end{cases}, \qquad (45)$$

where the non-dimensional parameters are introduced $\overline{f}_i = \dfrac{f_i \ell}{12H}$, $\overline{c}_i = \dfrac{c_i}{12H}$, $I_\psi = \dfrac{m\ell^2}{12H}$, $I_\varphi = \dfrac{I}{12H}$, $\tilde{K}_\psi = \dfrac{K_\psi \ell^2}{12H}$ and $\tilde{K}_\varphi = \dfrac{K_\varphi}{12H}$. The propagation of elastic waves along the 1D system is analyzed under the customary hypothesis of harmonic waves $\psi_i(t) = \overline{\psi} \exp\left[ j(kx_i - \omega t) \right]$ and $\varphi_i(t) = \overline{\varphi} \exp\left[ j(kx_i - \omega t) \right]$, and the dispersion functions are obtained as solution of the eigenvalue problem

$$\left( \mathbf{H}_{Lag} - \omega^2 \mathbf{I}_{Lag} \right)\upsilon = \left( \begin{bmatrix} 2\left[1 - \cos(k\ell)\right] + \tilde{K}_\psi & j\sin(k\ell) \\ -j\sin(k\ell) & \frac{1}{3}\left[2 + \cos(k\ell)\right] + \tilde{K}_\varphi \end{bmatrix} - \omega^2 \begin{bmatrix} I_\psi & 0 \\ 0 & I_\varphi \end{bmatrix} \right) \begin{Bmatrix} \overline{\psi} \\ \overline{\varphi} \end{Bmatrix} = \mathbf{0}.$$

(46)

being $\mathbf{H}_{Lag}$ an Hermitian matrix, $\mathbf{I}_{Lag}$ a diagonal matrix and $\upsilon$ is the polarization vector. In the Bloch spectrum the two solutions of (46) may represent either (a) an acoustic and an optical branch or (b) two optical branches, depending on the stiffness of the nodal elastic restraints. In fact, for the long wavelength limit $k\ell \to 0$, the angular frequency are $\omega_{0\psi} = \sqrt{\dfrac{\tilde{K}_\psi}{I_\psi}}$ and $\omega_{0\varphi} = \sqrt{\dfrac{1+\tilde{K}_\varphi}{I_\varphi}}$, from which it turns out that an acoustic branch may be obtained only if $\tilde{K}_\psi = 0$.

*Enhanced continualization via first order regularization approach*

To obtain an equivalent continuum, a macro-displacement field $\Psi(x_i,t)$ and a macro-rotation field $\Phi(x_i,t)$ are considered in analogy to the approach presented in Section 2. Once assumed the up-scaling law based on (10) and introduced the shift operator, the down-scaling law is obtained

$$\psi_i(t) = \dfrac{2\ell D}{\exp(\ell D) - \exp(-\ell D)} \Psi(x,t)\bigg|_{x_i}, \quad \varphi_i(t) = \dfrac{2\ell D}{\exp(\ell D) - \exp(-\ell D)} \Phi(x,t)\bigg|_{x_i}. \qquad (47)$$

According to the procedure involving the shift operator presented in Sections 1 and 2, the system of pseudo-differential equation is obtained

$$\begin{cases} \left\{ \dfrac{2\left[\exp(\ell D) - (2+\tilde{K}_\psi) + \exp(-\ell D)\right]}{\left[\exp(\ell D) - \exp(-\ell D)\right]} \ell D \right\} \Psi - \ell D \Phi + \bar{f} = \left\{ \dfrac{2 I_\psi}{\left[\exp(\ell D) - \exp(-\ell D)\right]} \ell D \right\} \ddot{\Psi} \\ \ell D \Psi - \left\{ \dfrac{\left[\exp(\ell D) + (4+6\tilde{K}_\varphi) + \exp(-\ell D)\right]}{3\left[\exp(\ell D) - \exp(-\ell D)\right]} \ell D \right\} \Phi + \bar{c} = \left\{ \dfrac{2 I_\varphi}{\left[\exp(\ell D) - \exp(-\ell D)\right]} \ell D \right\} \ddot{\Phi} \end{cases}, \qquad (48)$$

Through an expansion into McLaurin series in $D$ of the pseudo-differential operators in the braces retaining the terms up to the second order, the field equations of the homogenized model are obtained

$$\begin{cases} -\tilde{K}_\psi \Psi + \left(1 + \dfrac{1}{6}\tilde{K}_\psi\right) \ell^2 \dfrac{\partial^2 \Psi}{\partial x^2} - \ell \dfrac{\partial \Phi}{\partial x} + \bar{f}\ell = I_\psi \left( \ddot{\Psi} - \dfrac{\ell^2}{6} \dfrac{\partial^2 \ddot{\Psi}}{\partial x^2} \right) \\ \ell \dfrac{\partial \Psi}{\partial x} - (1+\tilde{K}_\varphi) \Phi + \dfrac{1}{6}\tilde{K}_\varphi \ell^2 \dfrac{\partial^2 \Phi}{\partial x^2} + \bar{c}\ell = I_\varphi \left( \ddot{\Phi} - \dfrac{\ell^2}{6} \dfrac{\partial^2 \ddot{\Phi}}{\partial x^2} \right) \end{cases}, \qquad (49)$$

It is worth to note that both the elastic potential energy density and the kinetic energy density in the Lagrangian density function associated to (49) turn out to be positive definite

$$\Pi_e = \dfrac{1}{2}\left[ \left(\ell \dfrac{\partial \Psi}{\partial x} - \Phi\right)^2 + \tilde{K}_\psi \Psi^2 + \dfrac{1}{6}\tilde{K}_\psi \ell^2 \left(\dfrac{\partial \Psi}{\partial x}\right)^2 + \tilde{K}_\varphi \Phi^2 + \dfrac{1}{6}\tilde{K}_\varphi \ell^2 \left(\dfrac{\partial \Phi}{\partial x}\right)^2 \right], \qquad (50)$$

$$T = I_\psi \left[ \dot\Psi^2 + \frac{1}{6}\ell^2 \left(\frac{\partial \dot\Psi}{\partial x}\right)^2 \right] + I_\varphi \left[ \dot\Phi^2 + \frac{1}{6}\ell^2 \left(\frac{\partial \dot\Phi}{\partial x}\right)^2 \right]. \qquad (51)$$

Retaining fourth order terms in the McLaurin series, the system of differential equations of the equivalent homogenized continuum is obtained

$$\begin{cases} -\tilde{K}_\psi \Psi + \left(1 + \frac{1}{6}\tilde{K}_\psi\right)\ell^2 \frac{\partial^2 \Psi}{\partial x^2} - \left(\frac{1}{12} + \frac{7}{360}\tilde{K}_\psi\right)\ell^4 \frac{\partial^4 \Psi}{\partial x^4} - \ell \frac{\partial \Phi}{\partial x} + \overline{f}\ell = I_\psi \left( \ddot\Psi - \frac{\ell^2}{6}\frac{\partial^2 \ddot\Psi}{\partial x^2} + \frac{7\ell^4}{360}\frac{\partial^4 \ddot\Psi}{\partial x^4} \right) \\ \ell \frac{\partial \Psi}{\partial x} - \left(1 + \tilde{K}_\varphi\right)\Phi + \frac{1}{6}\tilde{K}_\varphi \ell^2 \frac{\partial^2 \Phi}{\partial x^2} - \left(\frac{1}{180} + \frac{7}{360}\tilde{K}_\varphi\right)\ell^4 \frac{\partial^4 \Phi}{\partial x^4} + \overline{c}\ell = I_\varphi \left( \ddot\Phi - \frac{\ell^2}{6}\frac{\partial^2 \ddot\Phi}{\partial x^2} + \frac{7\ell^4}{360}\frac{\partial^4 \ddot\Phi}{\partial x^4} \right) \end{cases}, \qquad (52)$$

and the elastic potential energy density takes the positive definite form

$$\Pi_e = \frac{1}{2}\left[ \left(\ell \Psi_{,x} - \Phi\right)^2 + \tilde{K}_\psi \Psi^2 + \frac{1}{6}\tilde{K}_\psi \ell^2 \left(\frac{\partial \Psi}{\partial x}\right)^2 + \left(\frac{1}{12} + \frac{7}{360}\tilde{K}_\psi\right)\ell^4 \left(\frac{\partial^2 \Psi}{\partial x^2}\right)^2 + \right.$$
$$\left. + \tilde{K}_\varphi \Phi^2 + \frac{1}{6}\tilde{K}_\varphi \ell^2 \left(\frac{\partial \Phi}{\partial x}\right)^2 + \left(\frac{1}{180} + \frac{7}{360}\tilde{K}_\varphi\right)\ell^4 \left(\frac{\partial^2 \Phi}{\partial x^2}\right)^2 \right] \qquad (53)$$

as well as the kinetic energy density

$$T = I_\psi \left[ \dot\Psi^2 + \frac{1}{6}\ell^2 \left(\frac{\partial \dot\Psi}{\partial x}\right)^2 + \frac{7}{360}\ell^4 \left(\frac{\partial^2 \dot\Psi}{\partial x^2}\right)^2 \right] + I_\varphi \left[ \dot\Phi^2 + \frac{1}{6}\ell^2 \left(\frac{\partial \dot\Phi}{\partial x}\right)^2 + \frac{7}{360}\ell^4 \left(\frac{\partial^2 \dot\Phi}{\partial x^2}\right)^2 \right]. \qquad (54)$$

The plane wave propagation in the continuum models is characterized by the following eigenproblem

$$\left(\mathbf{H}_{Hom} - \omega^2 \mathbf{I}_{Hom}\right)\Upsilon = \mathbf{0}, \qquad (55)$$

being $\mathbf{H}_{Hom}$ an Hermitian matrix, $\mathbf{I}_{Hom}$ a diagonal matrix and $\Upsilon = \{\overline\Psi \ \overline\Phi\}^T$ is the polarization vector.

In the following two representative cases of fairly general systems are considered, which are characterized by a distinctive static and dynamic behavior.

*a. Transverse behavior of a rectangular beam lattice*

Let consider now the rectangular beam lattice with ligaments of length $\ell$ and $b$ and equal sections shown in Figure 7.a. Specific motions of the lattice are assumed, which are characterized by vanishing horizontal nodal displacements and equal vertical displacement and rotation of all the nodes located along a vertical line of ligaments, the generic one denoted as *i*. This kinematical assumption implies the uniformity of the forces $f_i$ and couples $c_i$ applied to all the nodes along the *i*-th vertical ligaments. From these hypotheses the motion of the lattice may be expressed in a mono-dimensional formulation and hence it may be analysed through the simplified model represented in Figure 7.b. In particular, the rollers are motivated by the vanishing of the curvature at the middle height of the vertical ligaments in Figure 7.a.

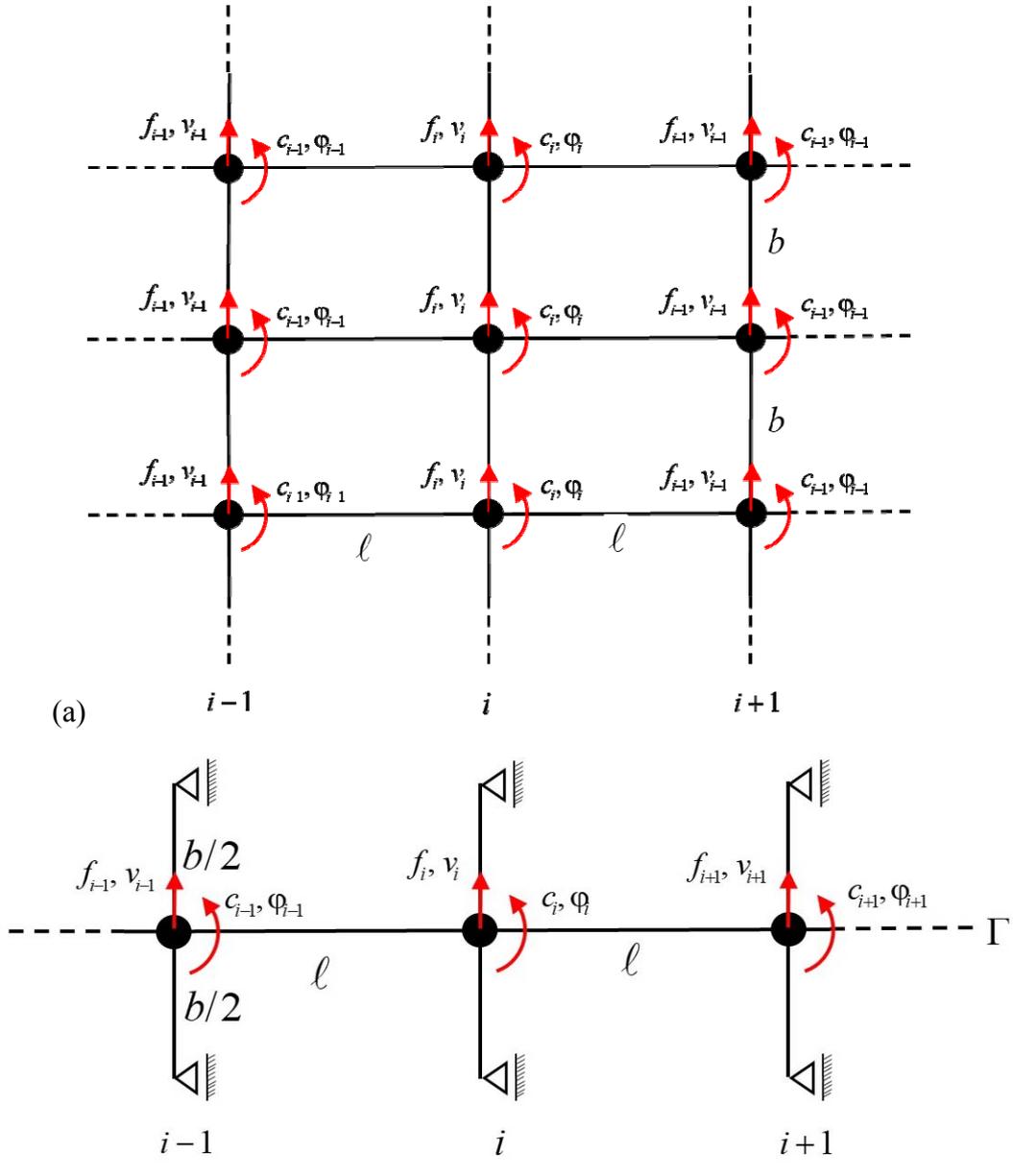

Figure 7. (a) rectangular lattice and assumed generalized nodal displacements and applied forces; (b) simplified mono-dimensional model.

The transverse deformation of the lattice in case of vanishing generalized nodal forces $\left(\bar{f}_i = \bar{c}_i = 0\right)$ may be analyzed through the equation of motion (45) with $K_\psi = 0$, namely $\tilde{K}_\psi = 0$. In this case, the equilibrium configurations of the lattice are obtained as solution of the difference equation (45) and are written in the form

$$\begin{cases} \psi_i = C_1 + C_2 i + C_3 \left(1 + 6\tilde{K}_\varphi + 2\sqrt{9\tilde{K}_\varphi^2 + 3\tilde{K}_\varphi}\right)^i + C_4 \left(1 + 6\tilde{K}_\varphi - 2\sqrt{9\tilde{K}_\varphi^2 + 3\tilde{K}_\varphi}\right)^i \\ \varphi_i = \left(1 + \tilde{K}_\varphi\right) C_2 - 2\frac{\sqrt{3}\sqrt{3\tilde{K}_\varphi^2 + \tilde{K}_\varphi}}{1 + 3\tilde{K}_\varphi} C_3 \left(1 + 6\tilde{K}_\varphi - 2\sqrt{9\tilde{K}_\varphi^2 + 3\tilde{K}_\varphi}\right)^i + \\ \qquad + 2\frac{\sqrt{3}\sqrt{3\tilde{K}_\varphi^2 + \tilde{K}_\varphi}}{1 + 3\tilde{K}_\varphi} C_4 \left(1 + 6\tilde{K}_\varphi + 2\sqrt{9\tilde{K}_\varphi^2 + 3\tilde{K}_\varphi}\right)^i \end{cases}, \quad (56)$$

(see Kelley and Peterson, 2001) with the constant $C_h$ to be obtained by the boundary conditions on displacements and rotations at the end nodes. Being positive the bases of the powers in equations (56) independently on $\tilde{K}_\varphi$, the configurations of this model are characterized by transverse displacements and rotations varying without oscillating sign reversal of the generalized displacements. In analogy to the solution of the discrete problem of Section 3, the solution (56) is characterized by a combination of linear and exponential terms, the latter ones providing boundary layers.

In case of square lattice $b = \ell$ one obtain $K_\varphi = 12H$, with $\tilde{K}_\varphi = 1$, and the solution takes the form

$$\begin{cases} \psi_i = C_1 + C_2 i + C_3 \left(7 - 4\sqrt{3}\right)^i + C_4 \left(7 + 4\sqrt{3}\right)^i \\ \varphi_i = 2C_2 - \sqrt{3} C_3 \left(7 - 4\sqrt{3}\right)^i + \sqrt{3} C_4 \left(7 + 4\sqrt{3}\right)^i \end{cases}. \quad (57)$$

To appreciate the approximation of the homogenized models, a first comparison with the solution of the Lagrangian model is carried out in the static field. Let consider a square beam lattice consisting of $n = 11$ ligaments in the direction along $\Gamma$ and an unbounded number of ligaments in the orthogonal direction. Let consider the homogeneous problem with vanishing rotation prescribed at the end nodes, namely $\varphi_0 = 0$ and $\varphi_{11} = 0$, while the non-dimensional transverse displacements are prescribed $\psi_0 = 0$ and $\psi_{11} = 10^{-2}$. In the diagrams of Figure 8 a comparison is given in terms of transverse displacement and rotation for the Lagrangian model with the corresponding results obtained by the proposed $2^{nd}$, $4^{th}$ and $6^{th}$ order homogenized models, i.e. by solving the governing equations obtained retaining $2^{nd}$, $4^{th}$ and $6^{th}$ order terms, respectively, in the McLaurin expansion of the pseudo-differential equations system (48). In addition, appropriate boundary conditions for the homogenized model, obtained via $r^{th}$ order continualization, are considered and take the form $\Psi(x=0)=0$, $\Psi(x=n\ell)=10^{-2}$, and $\Phi(x=0)=0$, $\Phi(x=n\ell)=0$, $\left.\dfrac{d^h\Psi}{dx^h}\right|_{x=0,n\ell}=\left.\dfrac{d^h\Phi}{dx^h}\right|_{x=0,n\ell}=0$, with $h=1,..,r/2$. From the diagrams in Figure 8 it may be observed a very good accuracy between the generalized macro-displacement field components $\Psi(x)$ and $\Phi(x)$ and the nodal solutions $\psi_i$ and $\varphi_i$ of the Lagrangian model. Indeed, the homogenised models are able to describe with good accuracy the boundary layer effects occurring in the discrete model. It is worth noting that this accuracy tends to increase as the order of continualization increases.

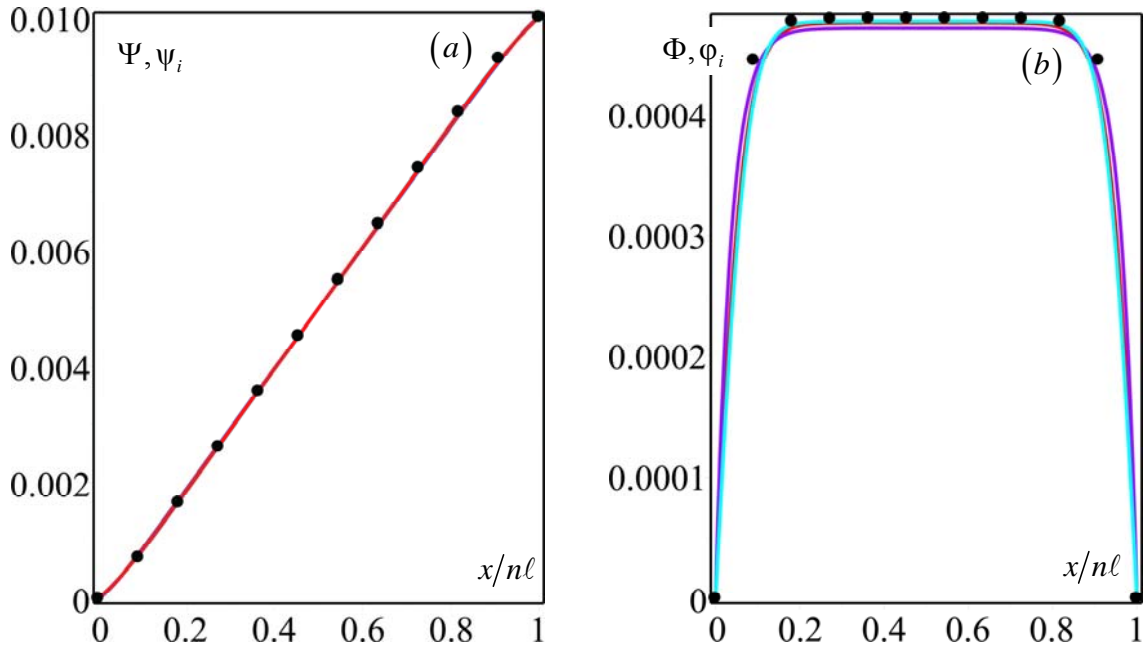

Figure 8. Transverse displacements (a) and rotations (b) in the Lagrangian model and from the continuum model. Comparison of different models: Lagrangian model (black dots); proposed $2^{nd}$ order continualization (violet); proposed $4^{th}$ order continualization (red); proposed $6^{th}$ order continualization (cyan).

A second comparison deals with the dispersion functions by the discrete model and those by the enhanced homogenized model; two values of the ratio between the translational and rotational inertia $\eta = \dfrac{I_\psi}{I_\varphi} = (10, 30)$ are considered to represent different design possibilities. Being $\tilde{K}_\psi = 0$, an acoustic and an optical branch are obtained by solving the eigenvalue problem (46), the latter departing from the critical frequency $\omega_{0\varphi} = \sqrt{\dfrac{1 + \tilde{K}_\varphi}{I_\varphi}}$, with group velocity vanishing and from where an optical branch departs. In the diagrams of Figure 9, the enhanced homogenized models are shown to provide a very good simulation of the Floquet-Bloch spectra of the Lagrangian model, with a decreasing band gap amplitude when decreasing the ratio $\eta$ between the translational and rotational inertia of the nodes. Moreover, increasing the continualization order, the dispersion functions of the derived enhanced continuum tend towards the actual corresponding branches.

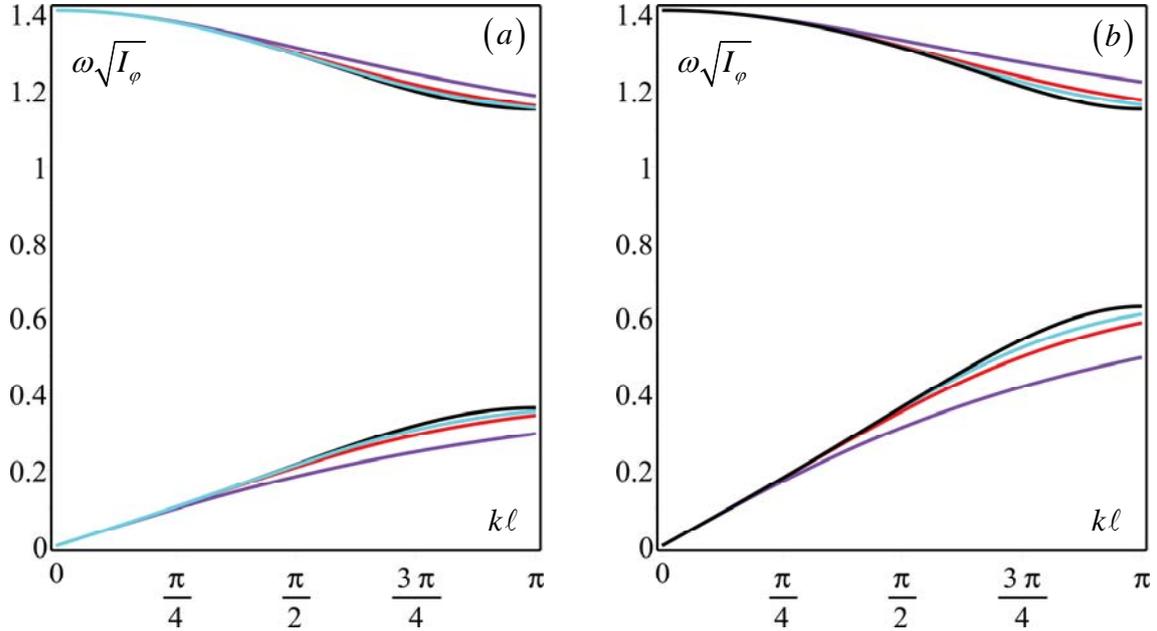

Figure 9. Dispersion functions for varying the mass ratio (a) $\eta = 30$, (b) $\eta = 10$: comparison among the Lagrangian model (black line) and the proposed continuum models: 2$^{nd}$ order continualization (violet); 4$^{th}$ order continualization (red); 6$^{th}$ order continualization (cyan).

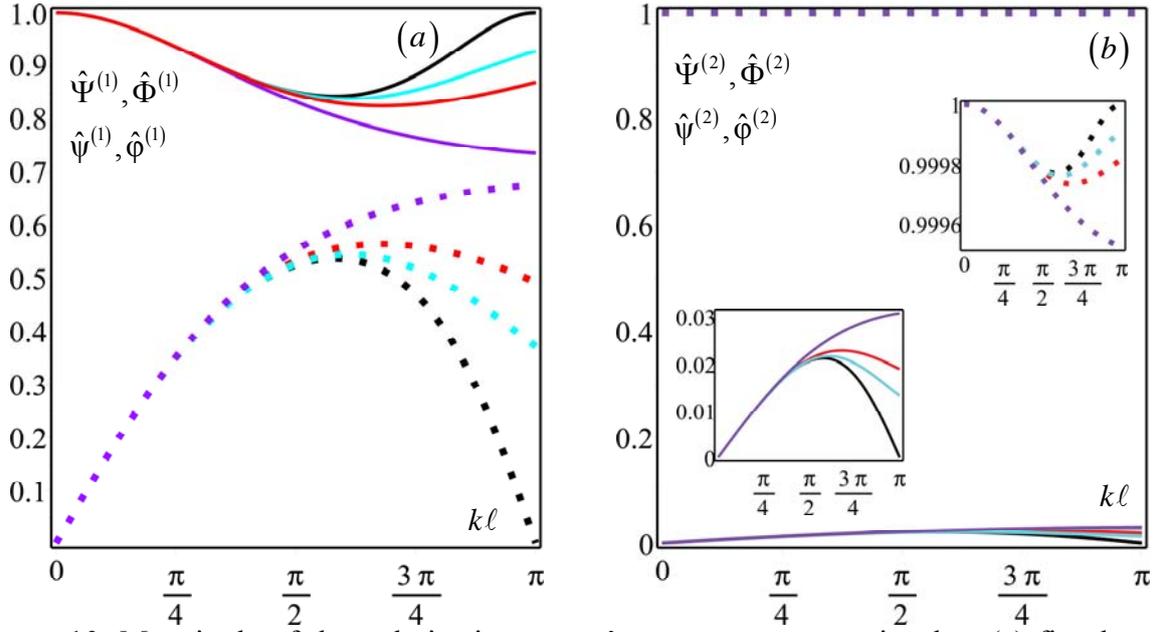

Figure 10. Magnitude of the polarization vector's components associated to (a) first branch and (b) second branch of frequency spectrum for mass ratios $\eta = 30$: comparison among the Lagrangian model (black line) and the proposed continuum models. First and second components of polarization vector in continuous and dot lines, respectively. Proposed 2$^{nd}$ order continualization (violet); proposed 4$^{th}$ order continualization (red); proposed 6$^{th}$ order continualization (cyan).

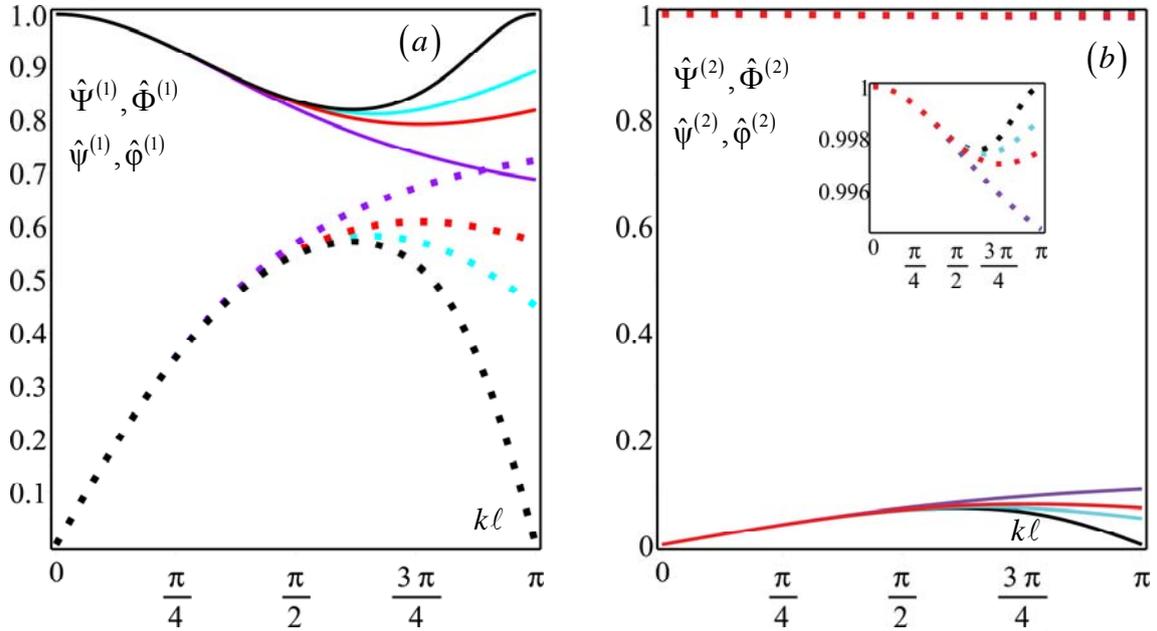

Figure 11. Magnitude of the polarization vector's components associated to (a) first branch and (b) second branch of frequency spectrum for mass ratios $\eta = 10$: comparison among the Lagrangian model (black line) and the proposed continuum models. First and second components of polarization vector in continuous and dot lines, respectively. Proposed 2$^{nd}$ order continualization (violet); proposed 4$^{th}$ order continualization (red); proposed 6$^{th}$ order continualization (cyan).

The description of the harmonic wave propagation is complemented with the analysis of the polarizations vectors. In particular, the polarization vectors $\mathbf{\upsilon} = \{\bar{\psi} \quad \bar{\varphi}\}^T$ and $\Upsilon = \{\bar{\Psi} \quad \bar{\Phi}\}^T$ associated to the Lagrangian and continuum models, respectively, are self-normalized by imposing the conditions $\mathbf{\upsilon} \cdot \mathbf{\upsilon}^* = 1$, $\Upsilon \cdot \Upsilon^* = 1$, where the superscript $^*$ stands for complex conjugate. The magnitude of the self-normalized polarization vector's components for the Lagrangian model $\hat{\psi} = \|\bar{\psi}\| = \sqrt{\bar{\psi}\bar{\psi}^*}$, $\hat{\varphi} = \|\bar{\varphi}\| = \sqrt{\bar{\varphi}\bar{\varphi}^*}$ and for the continuum model $\hat{\Psi} = \|\bar{\Psi}\| = \sqrt{\bar{\Psi}\bar{\Psi}^*}$, $\hat{\Phi} = \|\bar{\Phi}\| = \sqrt{\bar{\Phi}\bar{\Phi}^*}$ are determined in terms of the mechanical parameters and of the wave number. In the diagrams of Figures 10 and 11, these magnitudes are shown in terms of dimensionless wave number for mass ratios $\eta = 30$ and $\eta = 10$, respectively. In details, in sub-figures 10a and 11a the magnitudes of the components of the polarization vectors associated to the acoustic branch of the spectrum are reported, where a perfectly polarized displacement waveform is detected (namely, the eigenvector is mainly characterized by the first component) in the limit cases of long and short wavelengths, i.e. $k\ell = 0$ and $k\ell = \pi$. Moreover, for intermediate values of the dimensionless wave number the hybridization phenomena of the waveform components occur (in the sense that the waveform presents a transition from perfectly polarized waves to waves characterized by both the components of the eigenvector). The magnitudes of the polarization vector's components associated to the optical branch are shown in sub-figures 10b and 11b. Note that, a nearly polarized rotation waveform is detected to the except for limit cases $k\ell = 0$, $k\ell = \pi$ where a perfectly polarized rotation waveform occur. Finally, a good accuracy between the results obtained by the continuum models and those determined by Lagrangian model are observed into the range $0 \leq k\ell \leq \pi/2$. This accuracy tends to increase as the order of continualization increases.

*b. Continuous periodic beam on elastic supports*

The second system here considered concerns a continuous beam with elastic supports made up of beams pinned at the ends and having the same bending stiffness *EJ* and length *b* as shown in Figure 12. The lateral stiffness of the equivalent elastic spring is $K_\psi = 48 \frac{EJ}{b^3}$, while the rotational stiffness is vanishing $K_\varphi = 0$. The transverse deformation of the periodic beam in case of vanishing generalized nodal forces $\left(\overline{f}_i = \overline{c}_i = 0\right)$ may be analyzed through the equation of motion (45) with $\tilde{K}_\varphi = 0$ and $\tilde{K}_\psi = 4\left(\dfrac{\ell}{b}\right)^3$. This parameter has an influence on the quality of the solution of the system of difference equations. In fact, as explained in Appendix E, for $0 < \tilde{K}_\psi < 12$ the discrete model is characterized by a solution that harmonically varies along the beam with wavelength that depends on $\tilde{K}_\psi$. For increasing the transverse stiffness of the supports, namely increasing $\tilde{K}_\psi$, the wavelength decreases and for $\tilde{K}_\psi \geq 12$ a transition to an oscillating response with change of sign of the generalized displacements from one node to the adjacent one is obtained.

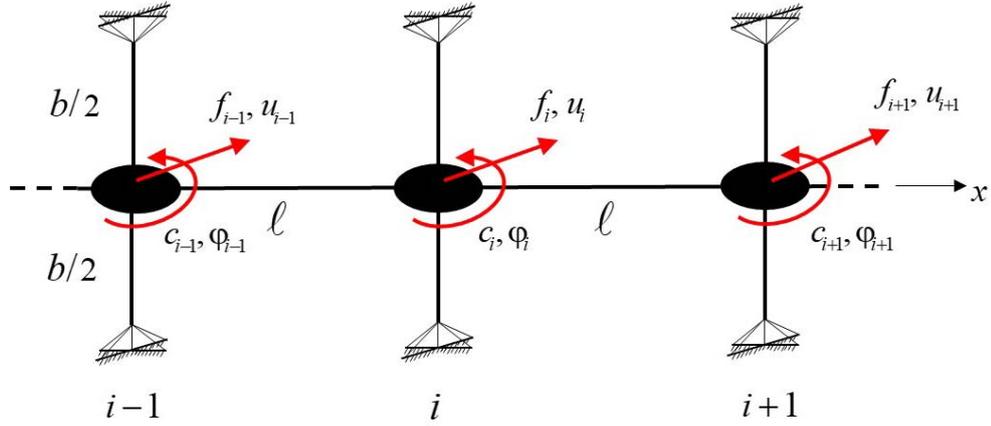

Figure 12. Periodic continuous beam with elastic supports.

This behavior is explained in the examples concerning a continuous beam consisting of $n = 11$ ligaments. The homogeneous problem is defined by prescribing vanishing rotations at the end nodes, namely $\varphi_0 = 0$ and $\varphi_{11} = 0$, while the non-dimensional transverse displacements are prescribed $\psi_0 = 0$ and $\psi_{11} = 10^{-1}$. In the diagrams of Figure 13 a comparison of transverse displacement and rotation of the nodes of the Lagrangian model with the corresponding results obtained by the proposed 4$^{th}$, 6$^{th}$ and 8$^{th}$ order homogenized models is given for the case of soft supports $\tilde{K}_\psi = 1/50$. Here, appropriate boundary conditions for homogenized model, obtained via $r^{th}$ order continualization, are considered and take the form $\Psi(x=0) = 0$, $\Psi(x=n\ell) = 10^{-2}$, and $\Phi(x=0) = 0$, $\Phi(x=n\ell) = 0$,

$\left.\dfrac{d^h \Psi}{dx^h}\right|_{x=0,n\ell} = \left.\dfrac{d^h \Phi}{dx^h}\right|_{x=0,n\ell} = 0$, with $h = 1,..,r/2$.

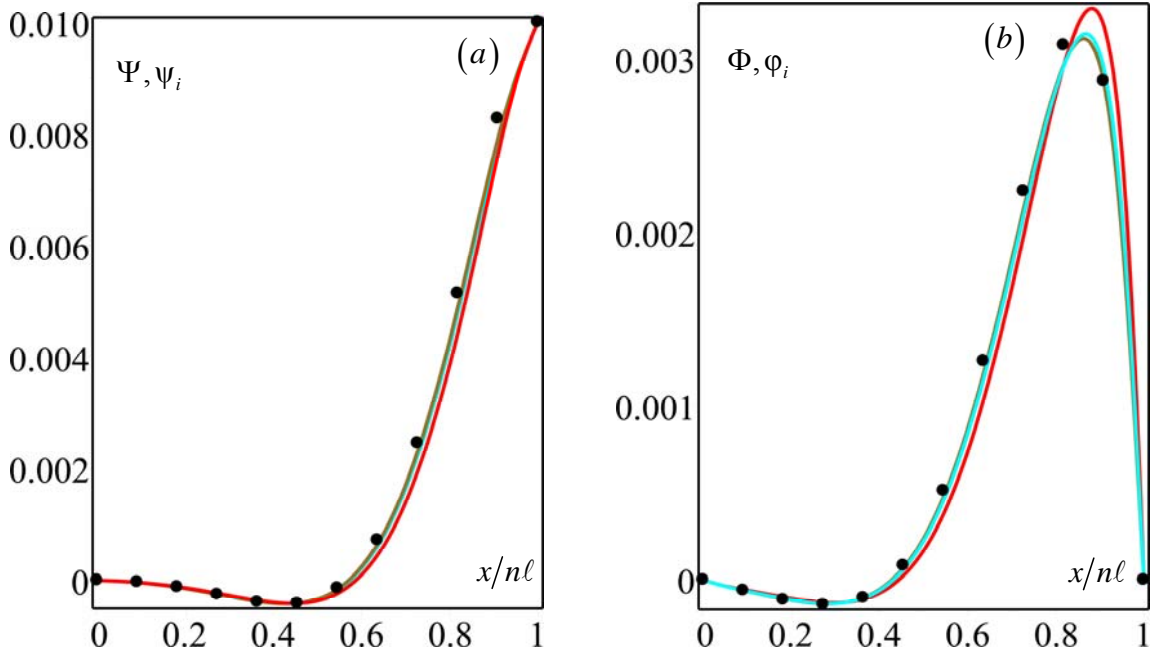

Figure 13. Transverse displacements (a) and rotations (b) in the Lagrangian model and from the continuum model for soft supports $\tilde{K}_\psi = 1/50$. Comparison of different models: Lagrangian model (black dots); proposed 4$^{th}$ order continualization (red); proposed 6$^{th}$ order continualization (cyan); proposed 8$^{th}$ order continualization (brown).

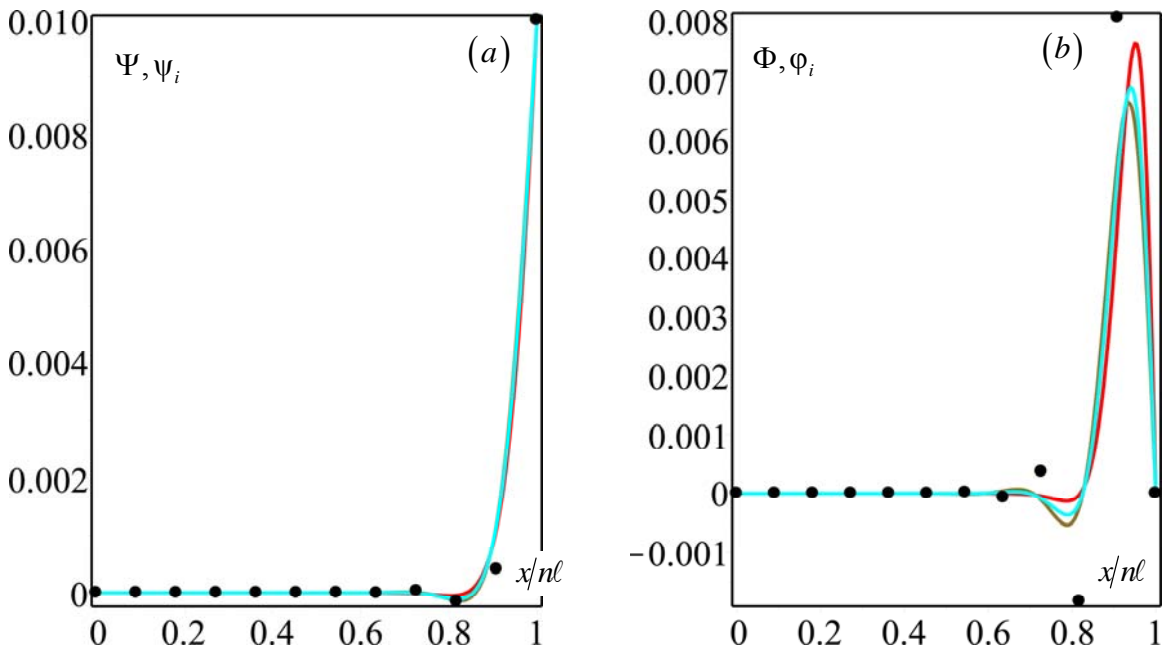

Figure 14. Transverse displacements (a) and rotations (b) in the Lagrangian model and from the continuum model for stiff supports $\tilde{K}_\psi = 20$. Comparison of different models: Lagrangian model (black dots); proposed 4$^{th}$ order continualization (red); proposed 6$^{th}$ order continualization (cyan); proposed 8$^{th}$ order continualization (brown).

From these diagrams it may be observed a very good accuracy between the generalized macro-displacement fields $\Psi(x)$ and $\Phi(x)$, respectively, of the continuum models obtained via enhanced homogenization here proposed and the nodal solutions $\psi_i$ and $\varphi_i$ resulting from the Lagrangian model. It is worth noting that this accuracy tends to increase as the order of continualization increases. Similarly, in the diagrams of Figure 14 a comparison for the case of stiff supports $\tilde{K}_\psi = 20$ is given. In this case it may be observed that the oscillating behavior of the discrete model in the boundary layer is simulated rather well by the enhanced continuum model here proposed, with a tendency to converge to the actual solution when increasing the order of the continuum model.

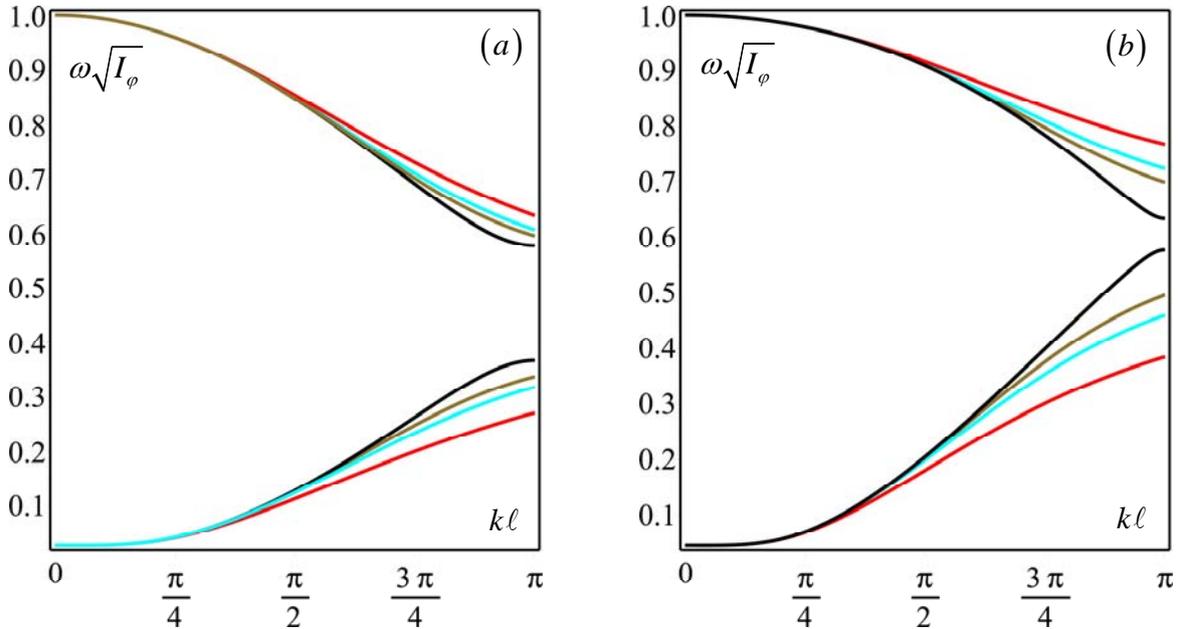

Figure 15. Dispersion functions for varying the mass ratio (a) $\eta = 30$, (b) $\eta = 10$ and for $\tilde{K}_\psi = 1/50$: comparison among the Lagrangian model (black line) and the proposed continuum models. Proposed $4^{th}$ order continualization (red); proposed $6^{th}$ order continualization (cyan); proposed $8^{th}$ order continualization (brown).

The evaluation of the homogenized model to simulate the acoustic behavior of the Lagrangian model has been carried out by considering two values of the ratio $\eta = I_\psi / I_\varphi = (10, 30)$ to represent several design possibilities. Being the rotational stiffness of the supports vanishing $\tilde{K}_\varphi = 0$, two optical branches are obtained by the eigenvalue problem (46). These two branches depart from corresponding critical frequencies $\omega_{0\psi} = \sqrt{\tilde{K}_\psi / I_\psi}$ and $\omega_{0\varphi} = \sqrt{1/I_\varphi}$, respectively. It is worth to note that while the second frequency only depends on the rotational inertia of the nodes, the former depends on the transverse stiffness of the supports and on the corresponding inertia. This circumstance suggests the possibility of a fine tuning on the optical branches also in designing guide-waves etc. Moreover, in case of $\tilde{K}_\psi \to 0$, the lower optical branch changes in an acoustic branch. In the diagrams of Figures 15 and 18 the Floquet-Bloch spectra obtained by the enhanced homogenised models are compared with those from the beam lattice for soft ($\tilde{K}_\psi = 1/50$) and stiff supports ($\tilde{K}_\psi = 20$). Although a good agreement is observed in the first case (see Figure 15), excellent results are obtained in case of stiff supports (see Figure 18). Also for this case one may observe that increasing the continualization order, the dispersion functions of the derived enhanced continuum tend towards the actual corresponding branches.

The description of the harmonic wave propagation is complemented with the analysis of the polarizations vectors. The magnitude of the self-normalized polarization vector's components for the Lagrangian model $\hat{\psi}$, $\hat{\varphi}$ and for the continuum model $\hat{\Psi}$, $\hat{\Phi}$ is evaluated in terms of the mechanical parameters and of the wave number. In the diagrams of Figures 16 and 17, these magnitudes are shown for soft supports ($\tilde{K}_\psi = 1/50$) in terms of dimensionless wave number for mass ratios $\eta = 30$ and $\eta = 10$, respectively. In particular, in sub-figures 16a and 17a the magnitudes of the polarization vector's components associated to the lowest frequency optical branch (first branch) of the spectrum are reported. It is possible to observe that a perfectly polarized displacement and/or rotation waveform is detected in the limit cases of long and short wavelengths, i.e. $k\ell = 0$ and $k\ell = \pi$. Moreover, for intermediate values of the dimensionless wave number the hybridization phenomena of the waveform's components occur. The magnitudes of the polarization vector's components associated to the highest frequency optical branch (second branch) are shown in sub-figures 16b and 17b. Note that, a nearly polarized rotation waveform (see sub-figure 16b) and hybridization phenomena of the waveform's components (see sub-figure 17b) are detected to the except for limit cases $k\ell = 0$, $k\ell = \pi$ where a perfectly polarized displacement and/or rotation waveform occur. Finally, a good accuracy between the results obtained by the continuum models and those determined by Lagrangian model are observed into the range $0 \le k\ell \le 3\pi/4$. This accuracy tends to increase as the order of continualization increases. Similarly, in the diagrams of Figures 19 and 20 a comparison for the case of stiff supports ($\tilde{K}_\psi = 20$) is given and qualitatively similar behaviors are detected.

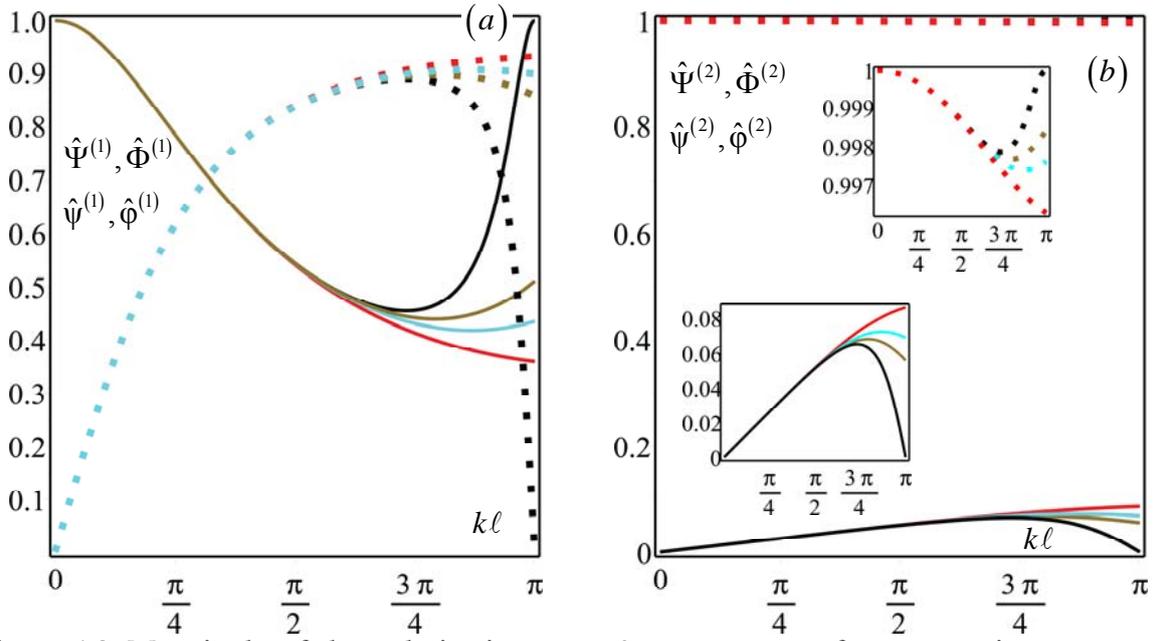

Figure 16. Magnitude of the polarization vector's components, for mass ratios $\eta = 30$ and $\tilde{K}_\psi = 1/50$ associated to (a) first branch and (b) second branch of frequency spectrum: comparison among the Lagrangian model (black line) and the proposed continuum models. First and second components of polarization vector in continuous and dot lines, respectively. Proposed 4$^{th}$ order continualization (red); proposed 6$^{th}$ order continualization (cyan); proposed 8$^{th}$ order continualization (brown).

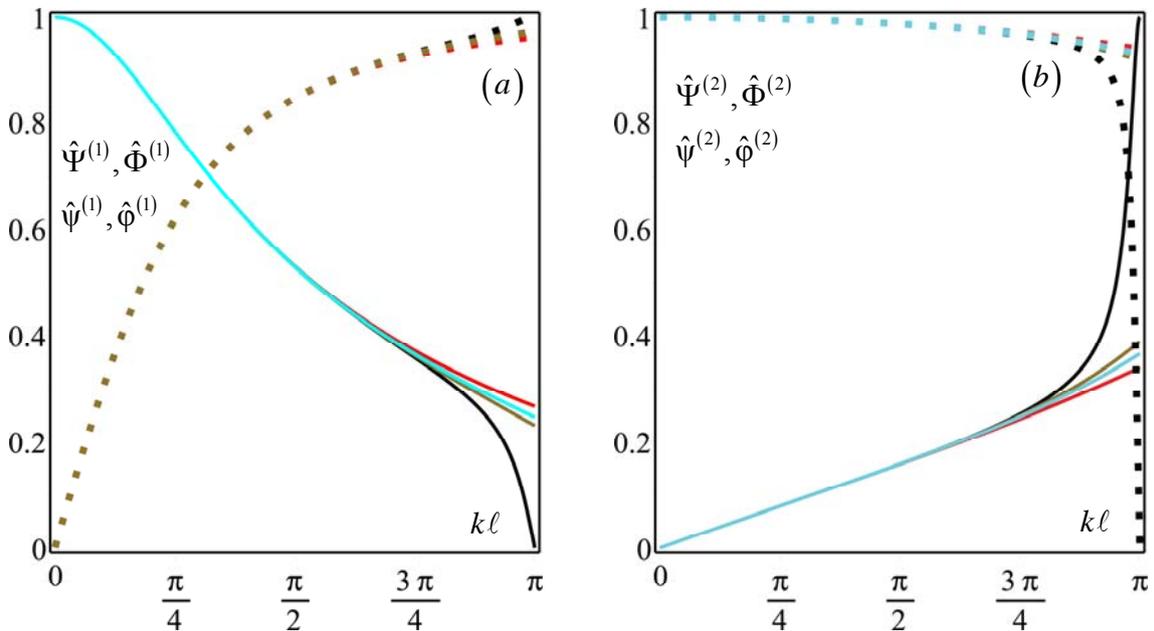

Figure 17. Magnitude of the polarization vector's components, for mass ratios $\eta = 10$ and $\tilde{K}_\psi = 1/50$ associated to (a) first branch and (b) second branch of frequency spectrum: comparison among the Lagrangian model (black line) and the proposed continuum models. First and second components of polarization vector in continuous and dot lines, respectively. Proposed 4$^{th}$ order continualization (red); proposed 6$^{th}$ order continualization (cyan); proposed 8$^{th}$ order continualization (brown).

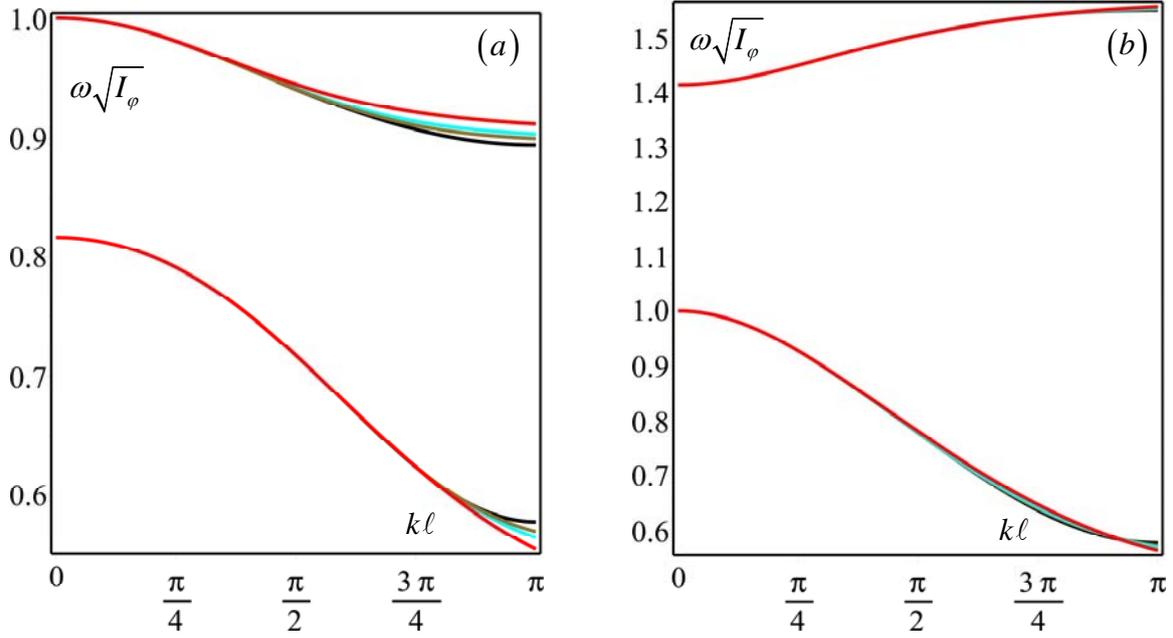

Figure 18. Dispersion functions for varying the mass ratio (a) $\eta = 30$, (b) $\eta = 10$ and for $\tilde{K}_\psi = 20$: comparison among the Lagrangian model (black line) and the proposed continuum models. Proposed 4$^{th}$ order continualization (red); proposed 6$^{th}$ order continualization (cyan); proposed 8$^{th}$ order continualization (brown).

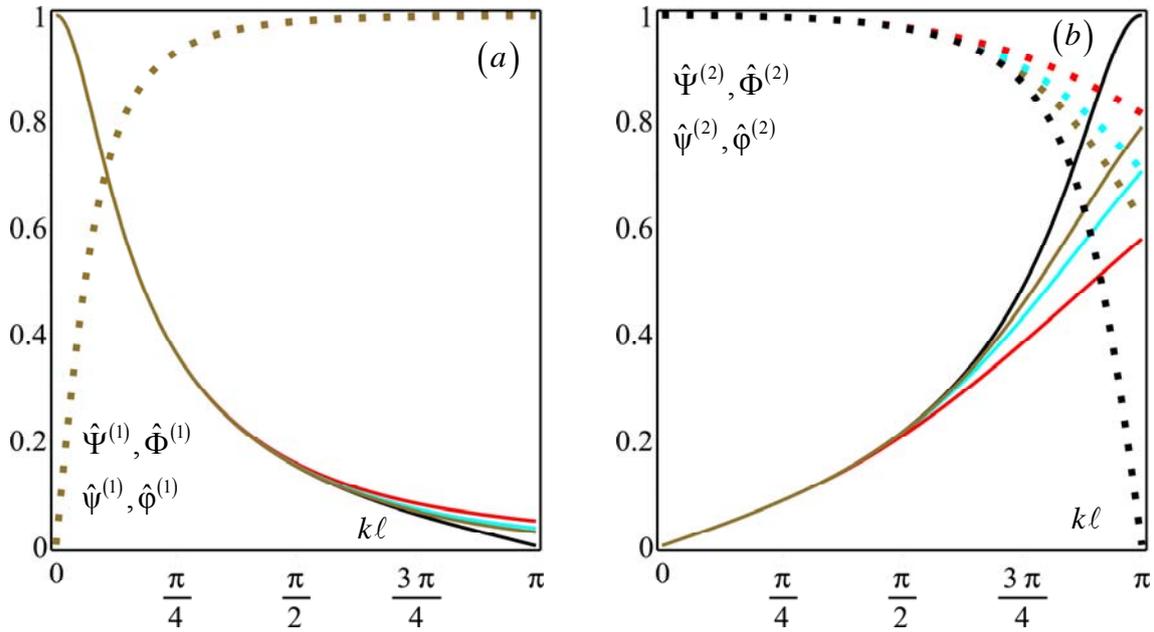

Figure 19. Magnitude of the polarization vector's components, for mass ratios $\eta = 30$ and $\tilde{K}_\psi = 20$ associated to (a) first branch and (b) second branch of frequency spectrum: comparison among the Lagrangian model (black line) and the proposed continuum models. First and second components of polarization vector in continuous and dot lines, respectively. Proposed 4$^{th}$ order continualization (red); proposed 6$^{th}$ order continualization (cyan); proposed 8$^{th}$ order continualization (brown).

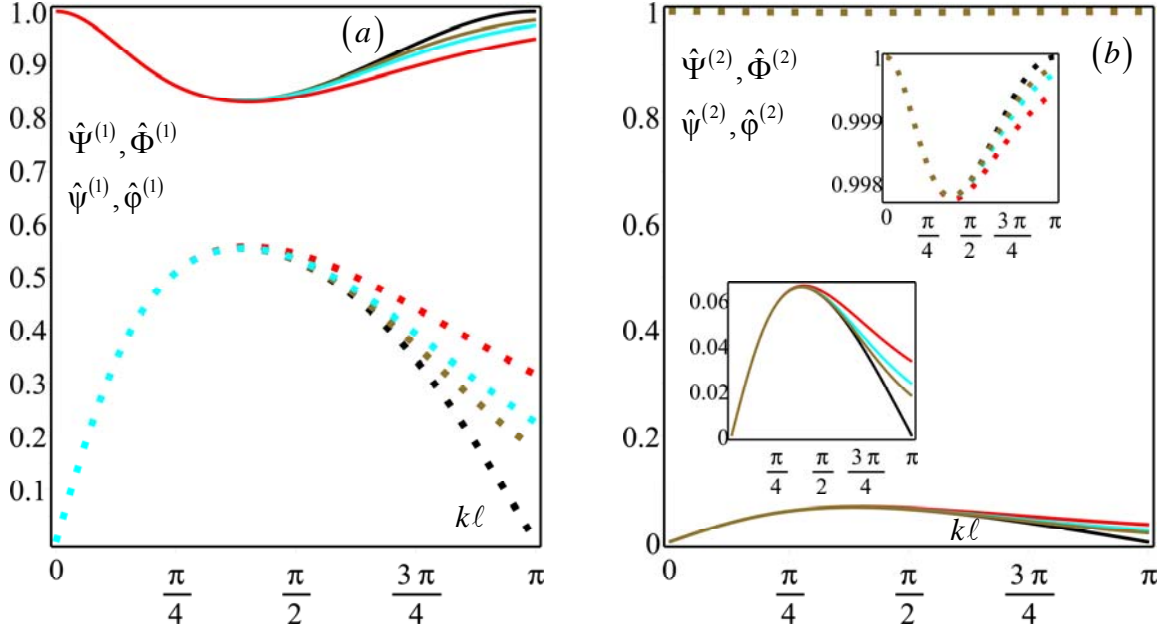

Figure 20. Magnitude of the polarization vector's components, for mass ratios $\eta = 10$ and $\tilde{K}_\psi = 20$ associated to (a) first branch and (b) second branch of frequency spectrum: comparison among the Lagrangian model (black line) and the proposed continuum models. First and second components of polarization vector in continuous and dot lines, respectively. Proposed 4$^{th}$ order continualization (red); proposed 6$^{th}$ order continualization (cyan); proposed 8$^{th}$ order continualization (brown).

## 6. Conclusions

The enhanced continualization approach proposed in this paper is aimed to overcome some drawbacks observed in the homogenization of beam lattices. While higher order micropolar models obtained through the standard continualization of beam lattices may provide good simulations of the acoustic response, on the other hand they are characterized by a non-positive definite elastic potential energy density. Conversely, micropolar homogenization approaches of beam lattice discrete models based on a first order down-scaling law coupled with an application of the macro-homogeneity condition appear not suitable to simulate the acoustic response of the lattice model. In the literature some attempts to circumvent such issues have been solved by the application of the Padé approximant to the pseudo-differential problem associated to the difference problem of the discrete model. Nevertheless, in this paper it has been shown that in some cases the approach based on the Padé approximant is unable to avoid non positive definiteness of the elastic potential of the homogenized continuum.

The approach here proposed, based on a transformation via a proper down-scaling law of the difference equation of motion of the discrete system into a pseudo-differential problem and a further McLaurin approximation to obtain a higher order differential problem, has been applied to some simple but significant 1D systems: the rod lattice, the beam lattice with node rotations and a 1D beam lattice model with generalized displacements. These systems present different static and dynamic characteristics. Nevertheless, the homogenized models obtained by the proposed enhanced continualization technique turn out to be energetically consistent and provide a good simulation of both the static response and of the acoustic spectrum of the original discrete models.

Higher order models may be obtained which are characterized by differential problems involving non-local inertia terms together with spatial high gradient terms. The simulation of the acoustic behaviour of the simple axial chain has shown a good performance if compared to the corresponding one from Padé approximation. The beam lattice with node rotations and displacement prevented exhibits, in the static regime, decaying oscillations of the nodal rotation in the boundary layer which is well simulated by the homogenized model obtained by the proposed approach. Similar good results are obtained in the simulation of the optical spectrum. It is worth to note that in this case the homogenized model obtained via Padé approximation turns out to be energetically non-consistent.

The analysis of the homogenized model derived from the beam lattice with transverse displacement and rotation of the nodes with elastic supports has shown that both the static and the dynamic response are strongly variable on the parameters of the Lagrangian model. Several different cases have been considered and good simulations have been obtained both in describing the static response to prescribed displacements at the end nodes and in representing the Floquet-Bloch spectrum and the polarization vectors for dimensionless wave number $k\ell \in \left[0, \frac{\pi}{2}\right]$. In consideration of these encouraging results, an implementation of the proposed enhanced continualization technique to 2D and 3D lattices will be presented in a forthcoming paper.

**Acknowledgements**

The authors acknowledge financial support of the (MURST) Italian Department for University and Scientific and Technological Research in the framework of the research MIUR Prin15 project 2015LYYXA8, Multi-scale mechanical models for the design and optimization of micro-structured smart materials and metamaterials, coordinated by prof. A. Corigliano.

**Appendix A – Formal definition of the pseudo-differential operator $P(D)$**

The equation of motion (4) written in terms of the pseudo-differential operator $P(D) = [\exp(\ell D) - 2 + \exp(-\ell D)]$ takes the form $P(D)\Psi + \bar{f} = I_\psi \ddot{\Psi}$. According to Maslov, 1976, and Shubin, 1987, the first term may be written as follows

$$P(D)\Psi(x,t) = \mathcal{F}^{-1}\left[P(j\xi)\mathcal{F}[\Psi(x,t)]\right] = \frac{1}{2\pi}\int_{-\infty}^{+\infty} P(j\xi)\mathcal{F}[\Psi(x,t)]e^{j\xi x}d\xi \quad \xi \in \mathbb{R}, \quad \text{(A.1)}$$

where $\mathcal{F}[\bullet] = \int_{-\infty}^{+\infty}(\bullet) e^{-j\xi x}dx$ and $\mathcal{F}^{-1}[\bullet] = \frac{1}{2\pi}\int_{-\infty}^{+\infty}(\bullet) e^{j\xi x}d\xi$ are the Fourier transform and the inverse Fourier transform, respectively, and $P(j\xi) = \exp(j\ell\xi) - 2 + \exp(-j\ell\xi)$ is an algebraic operator associated to the pseudo-differential operator $P(D)$.

**Appendix B – Definition of the pseudo-differential down-scaling law**

If an auxiliary displacement field $\tilde{\Psi}(x,t)$ satisfying the conditions $\tilde{\Psi}(x_i,t) = \psi_i(t)$ and $\tilde{\Psi}(x_{i\pm 1},t) = \psi_{i\pm 1}(t)$ is introduced, equation (11) may be written as follows

$$D\Psi(x,t) = \left[\frac{\exp(\ell D) - \exp(-\ell D)}{2\ell}\right]\tilde{\Psi}(x,t) = G(D)\tilde{\Psi}(x,t), \quad \text{(B.1)}$$

where the pseudo-differential operator $G(D)$ is defined. The Fourier transform of equation (B.1) may be rewritten in the form

$$j\xi\mathcal{F}[\Psi(x,t)] = G(j\xi)\mathcal{F}[\tilde{\Psi}(x,t)], \quad \text{(B.2)}$$

where $G(j\xi)$ is an algebraic operator associated to the pseudo-differential operator $G(D)$ and the Fourier transform of the auxiliary field $\tilde{\Psi}(x,t)$ turns out to be

$$\mathcal{F}\left[\tilde{\Psi}(x,t)\right] = \frac{j\xi}{G(j\xi)} \mathcal{F}\left[\Psi(x,t)\right] . \tag{B.3}$$

From the Fourier inverse transform of equation (B.3) it follows

$$\tilde{\Psi}(x,t) = \frac{1}{2\pi} \int_{-\infty}^{+\infty} \frac{j\xi}{G(j\xi)} \mathcal{F}\left[\Psi(x,t)\right] e^{j\xi x} d\xi \tag{B.4}$$

and remembering the definition (A.1), the equation (B.4) takes the form

$$\tilde{\Psi}(x,t) = Q(D)\Psi(x,t) \tag{B.5}$$

where the pseudo-differential operator $Q(D) = \left[\dfrac{D}{G(D)}\right]$ is introduced. Therefore, the pseudo-differential downscaling law (12) is obtained

$$\psi_i(t) = \tilde{\Psi}(x_i,t) = Q(D)\Psi(x,t)\big|_{x_i} . \tag{B.6}$$

**Appendix C – Continualization in integral non-local continuum**

The discrete governing equation of a generic 1-D lattice in which the generic node *i*, with mass *m*, is connected to the previous and subsequent *n* nodes via elastic ligaments, takes the following form

$$\sum_{p=-n}^{n} \alpha_p \psi_{i+p} + \bar{f}_i = I_\psi \ddot{\psi}_i , \tag{C.1}$$

being $\psi_i$ the non-dimensional axial displacement, $\bar{f}_i$ the non-dimensional axial force applied on the *i*-th node, $I_\psi$ and $\alpha_p$ the non dimensional inertial and constitutive parameters, respectively. In particular, for $n=1$ the 1-D lattice in which a generic node is connected only with the adjacent ones via two ligaments of length $\ell$, is recovered (see equation (1)). By performing a two-sided Zeta transform, the equation (C.1) takes the form

$$\mathcal{Z}\left[\sum_{p=-n}^{n} \alpha_p \psi_{i+p} + \bar{f}_i\right] = I_\psi \mathcal{Z}\left[\ddot{\psi}_i\right] , \tag{C.2}$$

where $\mathcal{Z}\left[h_i(t)\right] = \sum_{-\infty}^{+\infty} h_i(t) z^{-i} = \hat{h}(z,t)$ is the two-sided Zeta transform of the generic discrete variable $h_i(t)$ with $z \in \mathbb{C}$ and $i \in \mathbb{Z}$. Recalling the property $\mathcal{Z}\left[h_{i+p}(t)\right] = z^p \mathcal{Z}\left[h_i(t)\right] = z^p \hat{h}(z,t)$, the governing equation in the Z-space reads

$$\sum_{p=-n}^{n} \alpha_p z^p \hat{\psi}(z,t) + \hat{f}(z,t) = I_\psi \ddot{\hat{\psi}}(z,t). \tag{C.3}$$

By introducing the mapping $z = \exp(jk\ell)$, in which $k \in \mathbb{R}$ (that takes the meaning of the wave number), $j$ the imaginary unit and $\ell$ the distance between the adjacent nodes, the governing equation in Fourier space is expressed in the form

$$\sum_{p=-n}^{n} \alpha_p e^{jpk\ell} \hat{\psi}(z,t) + \hat{f}(z,t) = I_\psi \ddot{\hat{\psi}}(z,t). \tag{C.4}$$

Moreover, by performing an inverse Fourier transform one obtains

$$\mathcal{F}^{-1}\left[\sum_{p=-n}^{n} \alpha_p e^{jpk\ell} \hat{\psi}(z,t) + \hat{f}(z,t)\right] = \mathcal{F}^{-1}\left[I_\psi \ddot{\hat{\psi}}(z,t)\right], \tag{C.5}$$

which can be specialized in the equivalent form

$$\mathcal{F}^{-1}\left[\sum_{p=-n}^{n} \alpha_p e^{jpk\ell}\right] * \mathcal{F}^{-1}\left[\hat{\psi}(z,t)\right] + \mathcal{F}^{-1}\left[\hat{f}(z,t)\right] = I_\psi \mathcal{F}^{-1}\left[\ddot{\hat{\psi}}(z,t)\right], \tag{C.6}$$

where the symbol $*$ denotes the convolution product. The governing equation of integral non-local continuum is obtained by explicating the convolution product in equation (C.6), i.e.

$$\int_{-\infty}^{+\infty} \breve{g}(x-y) \breve{\psi}(y,t) dy + \breve{f}(x,t) = I_\psi \ddot{\breve{\psi}}(x,t), \tag{C.7}$$

being $\breve{g}(x-y)$ the kernel of non-local model, $\breve{\psi}(x,t)$ the macroscopic non-dimensional displacement, $\breve{f}(x,t)$ the macroscopic non-dimensional force, which are formally expressed as follows

$$\begin{aligned}
\breve{g}(x-y) &= \mathcal{F}^{-1}\left[\sum_{p=-n}^{n} \alpha_p e^{jpk\ell}\right]\bigg|_{x-y}, \\
\breve{\psi}(x,t) &= \mathcal{F}^{-1}\left[\hat{\psi}(k,t)\right], \\
\breve{f}(x,t) &= \mathcal{F}^{-1}\left[\hat{f}(k,t)\right].
\end{aligned} \tag{C.8}$$

Moreover, it may be demonstrated that the frequency spectrum obtained via integral non-local continuum coincides exactly with that obtained from the Lagrangian model. In fact, by performing a time and space Fourier transform of the governing equation (44) and considering null source term, the corresponding governing equation in the space of the angular frequencies and of the wave numbers takes the following form

$$\sum_{p=-n}^{n} \alpha_p e^{jpk\ell} \mathcal{F}_t\left[\hat{\psi}(k,t)\right] + \omega^2 I_\psi \mathcal{F}_t\left[\hat{\psi}(k,t)\right] = 0, \tag{C.9}$$

where $\mathcal{F}_t[\cdot] = \int_{-\infty}^{+\infty} (\cdot) e^{-j\omega t} dt$ is the time Fourier transform. From the equation (C.9), is possible to obtained the following dispersion relation

$$\sum_{p=-n}^{n} \alpha_p e^{jpk\ell} + \omega^2 I_\psi = 0, \tag{C.10}$$

that exactly coincides with the actual one obtained by the Lagrangian model through the discrete governing equation (C.1).

**Appendix D – Padé fourth order approximation for 1D beam lattice with node rotation**

Let consider now a fourth order Padé approximation of the l.h.s of the pseudo-differential equation (26)

$$-\frac{1}{6}\left[\exp(-\ell D) + 4 + \exp(\ell D)\right]\Phi \approx -\frac{1 + \frac{31}{252}(\ell D)^2 + \frac{113}{15120}(\ell D)^4}{1 - \frac{11}{252}(\ell D)^2 + \frac{13}{15120}(\ell D)^4}\Phi. \tag{D.1}$$

The resulting differential equation of motion of the equivalent homogenized continuum takes the form

$$-\frac{113}{15120}\ell^4 \frac{\partial^4 \Phi}{\partial x^4} - \frac{31}{252}\ell^2 \frac{\partial^2 \Phi}{\partial x^2} - \Phi + \overline{c}\ell = I_\varphi\left(\ddot{\Phi} - \frac{11}{252}\ell^2 \frac{\partial^2 \ddot{\Phi}}{\partial x^2} + \frac{13}{15120}\ell^4 \frac{\partial^4 \ddot{\Phi}}{\partial x^4}\right). \tag{D.2}$$

It is worth to note that the Lagrangian density function from which equation (40) is derived is

$$\mathcal{L} = \frac{1}{2\ell} I_\varphi \left[\dot{\Phi}^2 + \frac{11}{252}\ell^2 \left(\frac{\partial \dot{\Phi}}{\partial x}\right)^2 + \frac{13}{15120}\ell^4 \left(\frac{\partial^2 \dot{\Phi}}{\partial x^2}\right)^2\right] \\ - \frac{1}{2\ell}\left[\Phi^2 - \frac{\ell^2}{12}\left(\frac{\partial \dot{\Phi}}{\partial x}\right)^2 + \frac{113}{15120}\ell^4\left(\frac{\partial^2 \Phi}{\partial x^2}\right)^2\right] + \overline{c}\Phi, \tag{D.3}$$

and is characterized by a non-positive definite elastic potential energy density. As a consequence, also for this model an oscillating behavior with constant amplitude is obtained

$$\Phi(x,t) = \exp\left(A^- \frac{x}{\ell}\right)\left[G_1 \cos\left(A^+ \frac{x}{\ell}\right) + G_2 \cos\left(A^+ \frac{x}{\ell}\right)\right] +$$
$$+ \exp\left(-A^- \frac{x}{\ell}\right)\left[G_1 \cos\left(A^+ \frac{x}{\ell}\right) + G_2 \cos\left(A^+ \frac{x}{\ell}\right)\right],$$
(D.4)

with $A^\pm = \dfrac{\sqrt{678\sqrt{11865} \pm 52545}}{113}$ and $G_h$, $h=1,4$ constants to be determined from the boundary conditions. Finally, the dispersion function is

$$\omega = \sqrt{\frac{1 - \frac{31}{252}k^2\ell^2 + \frac{113}{15120}k^4\ell^4}{1 + \frac{11}{252}k^2\ell^2 + \frac{13}{15120}k^4\ell^4}} = \frac{1}{\sqrt{I_\varphi}}\left(1 - \frac{1}{12}k^2\ell^2 + \frac{1}{288}k^4\ell^4\right) + \mathcal{O}(k^6\ell^6). \quad \text{(D.5)}$$

**Appendix E – Solution of the static problem in Section 3.b**

Equilibrium motion ($\tilde{K}_\varphi = 0$) in case of vanishing generalized applied forces:

$$\begin{cases} \psi_{i-1} - (2 + \tilde{K}_\psi)\psi_i + \psi_{i+1} - \dfrac{1}{2}(\varphi_{i+1} - \varphi_{i-1}) = 0 \\ \dfrac{1}{2}(\psi_{i+1} - \psi_{i-1}) - \dfrac{1}{6}(\varphi_{i-1} + 4\varphi_i + \varphi_{i+1}) = 0 \end{cases}, \quad \text{(E.1)}$$

Once introduced the shift operator, the system (E.1) is solved according to Kelley and Peterson, 2001. The first step is the solution of the equation

$$\left\{-\frac{1}{6}(E^2 + 4E + 1)\left[E^2 - (2 + \tilde{K}_\psi)E + 1\right] + \frac{1}{2}(E^2 - 1)\right\}\psi_i = 0, \quad \text{(E.2)}$$

having roots:

$$E_1 = -\frac{1}{2}\tilde{K}_\psi + 1 + \frac{1}{2}S + \frac{\sqrt{2}}{2}\sqrt{\tilde{K}_\psi^2 - \tilde{K}_\psi S - 8\tilde{K}_\psi + 2S},$$
$$E_2 = -\frac{1}{2}\tilde{K}_\psi + 1 + \frac{1}{2}S - \frac{\sqrt{2}}{2}\sqrt{K_\psi^2 - \tilde{K}_\psi S - 8\tilde{K}_\psi + 2S},$$
$$E_3 = -\frac{1}{2}\tilde{K}_\psi + 1 - \frac{1}{2}S + \frac{\sqrt{2}}{2}\sqrt{\tilde{K}_\psi^2 - \tilde{K}_\psi S - 8\tilde{K}_\psi - 2S},$$
$$E_4 = -\frac{1}{2}\tilde{K}_\psi + 1 - \frac{1}{2}S - \frac{\sqrt{2}}{2}\sqrt{\tilde{K}_\psi^2 - \tilde{K}_\psi S - 8\tilde{K}_\psi - 2S},$$
(E.3)

being $S = \sqrt{\tilde{K}_\psi^2 - 12\tilde{K}_\psi}$. Two different kind of solutions may be obtained depending on the value of $\tilde{K}_\psi$.

**Case (a)** - $0 < \tilde{K}_\psi \leq 12$

In this case $S = \sqrt{\tilde{K}_\psi^2 - 12\tilde{K}_\psi} = j\sqrt{12\tilde{K}_\psi - \tilde{K}_\psi^2} \in \mathbb{C}$ and the solution takes the form:

$$\begin{cases} \psi_i = C_1 r_1 \cos(\theta_1 i) + C_2 r_1 \sin(\theta_1 i) + C_3 r_2 \cos(\theta_2 i) + C_4 r_2 \sin(\theta_2 i) \\ \varphi_i = G_1 r_1 \cos(\theta_1 i) + G_2 r_1 \sin(\theta_1 i) + G_3 r_2 \cos(\theta_2 i) + G_4 r_2 \sin(\theta_2 i) \end{cases} \quad \text{(E.4)}$$

with $r_h$ and $\theta_h$ the magnitude and the argument, respectively, of the complex numbers by (E.3), $C_k$ and $G_k$ being constants to be determined by the boundary conditions and satisfying the difference equation (E.1). It turns out that for $0 < \tilde{K}_\psi \leq 12$ the solution of the discrete problem is harmonic with wavelengths that depend on the parameter $\tilde{K}_\psi$.

**Case (b)** - $\tilde{K}_\psi > 12$

In this case $S \in \mathbb{R}$ and the solution takes the form:

$$\begin{cases} \psi_i = C_1 E_1^i + C_2 E_2^i + C_3 E_3^i + C_4 E_4^i \\ \varphi_i = G_1 E_1^i + G_2 E_2^i + G_3 E_3^i + G_4 E_4^i \end{cases} \quad \text{(E.5)}$$

where all the terms $E_h < 0$ and the constants $C_k$ and $G_k$ are to be determined by the boundary conditions and satisfying the difference equation (E.1). The solution (E.5) may be written in an alternative form

$$\begin{cases} \psi_i = (-1)^i \left( C_1 |E_1|^i + C_2 |E_2|^i + C_3 |E_3|^i + C_4 |E_4|^i \right) \\ \varphi_i = (-1)^i \left( G_1 |E_1|^i + G_2 |E_2|^i + G_3 |E_3|^i + G_4 |E_4|^i \right) \end{cases} \quad \text{(E.6)}$$

showing the oscillating property of the solution, with change of sign of the generalized displacement from one node to the adjacent one. For $\tilde{K}_\psi = 12$ one obtains from (E.3) $E_{1,2} = -5 - 2\sqrt{6}$ and $E_{3,4} = -5 + 2\sqrt{6} < 0$ with a transition from the harmonic regime (E.4) to the decaying oscillations (E.6).